\documentclass[sigconf, screen]{acmart}

\AtBeginDocument{%
  }

\setcopyright{none} 
\copyrightyear{2026}
\acmYear{2026}
\acmDOI{XXXXXXX.XXXXXXX}

\acmConference[MICRO 2026]{The 58th IEEE/ACM International Symposium on Microarchitecture}{October 31--November 04, 2026}{Athens, Greece}

\acmISBN{978-X-XXXX-XXXX-X/XX/XX}

\usepackage{amsmath,amssymb,amsfonts}
\usepackage{algorithm}
\usepackage{algorithmic}
\usepackage{graphicx}
\usepackage{multicol} 
\usepackage{textcomp}
\usepackage[table,dvipsnames]{xcolor}
\usepackage{fancyhdr}
\usepackage{listings}
\usepackage{hyperref}
\usepackage{pifont}
\usepackage{booktabs}
\usepackage{multirow}
\usepackage{caption}
\usepackage{enumitem}
\usepackage{xspace}

\newcommand*{\name}{CODA\xspace}
\newcommand*{\erase}[1]{}

\newcommand{\revision}[1]{{#1}}
\newcommand{\revsecref}[2]{\hyperref[#1]{Sec.~\ref*{#2}}}
\newcommand{\revfigref}[2]{\hyperref[#1]{Fig.~\ref*{#2}}}

\hypersetup{
    colorlinks=true,
    linkcolor=RoyalBlue,
    citecolor=BrickRed,
    urlcolor=RoyalBlue
}

\title{CODA: Algorithm-Hardware Co-design for Edge Video Diffusion\\via NMP-Enabled Compute-Cache Operator Disaggregation}

\author{Yuanpeng Zhang}
\affiliation{%
  \institution{Peking University}
  \city{Beijing}
  \country{China}}
\email{zyp\_cs@pku.edu.cn}

\author{YuXuan Wu}
\affiliation{%
  \institution{Peking University}
  \city{Beijing}
  \country{China}}
\email{2300011492@stu.pku.edu.cn}

\author{Yitong Xiao}
\affiliation{%
  \institution{Peking University}
  \city{Beijing}
  \country{China}}
\email{xiaoyt@stu.pku.edu.cn}

\author{Chenhao Xue}
\affiliation{%
  \institution{Peking University}
  \city{Beijing}
  \country{China}}
\email{xch927027@pku.edu.cn}

\author{Yi Ren}
\affiliation{%
  \institution{Peking University}
  \city{Beijing}
  \country{China}}
\email{yiren20@stu.pku.edu.cn}

\author{Cong Li}
\affiliation{%
  \institution{Peking University}
  \city{Beijing}
  \country{China}}
\email{leesou@pku.edu.cn}

\author{Yihan Yin}
\affiliation{%
  \institution{Peking University}
  \city{Beijing}
  \country{China}}
\email{yyhsess2021@stu.pku.edu.cn}

\author{Dimin Niu}
\affiliation{%
  \institution{Alibaba Group Inc.}
  \city{Hangzhou}
  \country{China}}
\email{dimin.niu@alibaba-inc.com}

\author{Guangyu Sun}
\affiliation{%
  \institution{Peking University}
  \city{Beijing}
  \country{China}}
\email{gsun@pku.edu.cn}

\renewcommand{\shortauthors}{Yuanpeng Zhang et al.}

\begin{document}

\hypersetup{pageanchor=true}
\setcounter{page}{1}


\begin{abstract}
Deploying Video Diffusion Models (VDMs) on edge devices is appealing for localized and privacy-preserving generation, but their iterative Transformer-based denoising remains too slow for practical local inference. Cross-Timestep Caching (CTC) has emerged as a promising direction for reducing redundant computation, reusing activations across adjacent denoising steps rather than modifying model weights, while largely preserving generation fidelity. However, on memory-constrained edge GPUs, CTC requires a massive cache footprint that quickly exceeds on-device VRAM and forces the cache into host memory. More fundamentally, cache operators remain tightly interleaved and chain-dependent with native compute operators, so naive near-memory offloading still incurs repeated PCIe exchanges for residual and fusion computations, turning cache reuse into a communication- and serialization-bound execution flow. We therefore propose CODA, an algorithm-hardware co-designed architecture centered on \underline{C}ompute-Cache \underline{O}perator \underline{D}is\underline{a}ggregation. CODA separates dense compute paths and memory-bound cache paths across the xPU and a lightweight DIMM-side near-memory engine, reorganizes fragmented cache activity into hardware-friendly coalesced segments, and exploits Classifier-Free Guidance (CFG) branch independence to overlap xPU compute with cache-side execution. Experiments show that CODA achieves up to 1.80$\times$ end-to-end speedup and 1.74$\times$ higher energy efficiency, while preserving competitive generation quality compared with a state-of-the-art caching algorithm. 
\end{abstract}
\keywords{Video Diffusion Model, Near-Memory Processing}
\maketitle

\section{Introduction}
\label{sec:intro}

Recent Video Diffusion Models (VDMs), exemplified by ByteDance Seedance 2.0~\cite{bytedance2025seedance} and OpenAI Sora 2~\cite{openai2025sora2}, have substantially improved the fidelity and physical consistency of video generation, making them an important foundation for digital content creation~\cite{xing_survey_2024,blattmann_stable_2023,kong2025hunyuanvideo, fan_vchitect-20_2025, yang_cogvideox_2025}. As deployment scenarios broaden and representative models continue to mature and become open-source~\cite{opensora-plan,yang_cogvideox_2025,kong2025hunyuanvideo,wan_wan_2025}, extending video generation from the cloud to edge devices such as GPU-equipped workstations becomes increasingly attractive for offline availability and privacy preservation~\cite{li_edge_2020,noauthor_mobilediffusion_nodate,zheng_diffusion_2025, kanipakam_privacy-preserving_2025}. However, their iterative denoising process over heavy transformers still incurs high inference latency. Existing model compression methods, such as quantization~\cite{efficientdm-quant,chu_qncd_2024} and pruning~\cite{laptop-pruning,li_optimizing_2025,becker_edit_2025}, can reduce computation, but often require costly fine-tuning for recovery~\cite{efficientdit-distillation,relic_bridging_2025,yao_timestep-aware_2024}.

\begin{figure*}[t]
  \centering
  \includegraphics[width=0.99\linewidth]{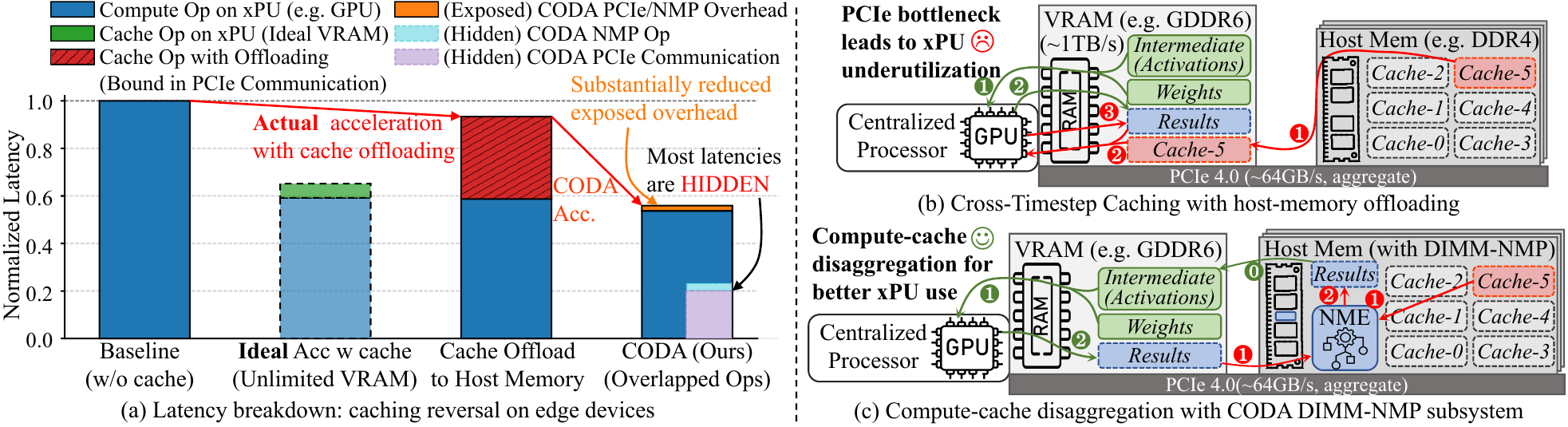}
\caption{Motivation and solution overview. (a) Host memory cache offloading can erase expected CTC gains on edge devices, while CODA substantially reduces end-to-end latency by hiding exposed cache-path overhead (xPU is NVIDIA RTX 4090 GPU). (b) Limited VRAM pushes overflow cache to host DRAM, making the cache path PCIe-bound and lowering xPU utilization. (c) CODA disaggregates cache operators to lightweight DIMM-NMP and overlaps cache-side execution with xPU compute.}
  \label{fig:teaser}
\end{figure*}

Prior studies~\cite{zhao_pab_2025,ToCa2025_localfix,selvaraju_fora_2024,fan_taocache_2025} have observed strong feature-map similarity between adjacent denoising timesteps, motivating \textbf{Cross-Timestep Caching (CTC)}~\cite{liu_reusing_2025,wu_invardiff_2025,liu_speca_2025}, which reuses intermediate features from previous timesteps (e.g., outputs of Attention or FFN modules) to bypass redundant computation. By exploiting cross-timestep similarity rather than perturbing model weights or activations, CTC remains orthogonal to conventional model compression and is particularly appealing for video generation. However, high-fidelity reuse requires caching a massive volume of intermediate features, and the cache footprint scales with model size, video resolution, and duration. For example, generating a 4-second 720p video with Open-Sora~\cite{zheng_open-sora_2024} requires more than 40~GB of cache storage, far beyond the typical VRAM capacity ($\sim$24~GB) of high-end edge GPUs. Existing software stacks therefore offload cross-timestep cache to host memory (DRAM)~\cite{zhao_pab_2025,cache-dit_2025,maurya_mlp-offload_2025,li_out_2026,liu_make_2025}. Since cache operators have extremely low arithmetic intensity, fetching massive cached data over PCIe quickly makes them bandwidth-bound, creating a severe performance wall, as shown in Figure~\ref{fig:teaser}(a) and Figure~\ref{fig:teaser}(b).

A natural idea is to offload these bandwidth-bound cache operators to a Near-Memory Processing (NMP) subsystem~\cite{kwon_tensordimm_2019,Lee_AxDIMM_2022,ke_recnmp_2020,park_attacc_2024}. However, this naive approach overlooks the strict serial dependencies in VDM execution. In existing CTC flows, native compute operators and cache operators remain tightly interleaved, and results produced by one side are often immediately required by the other. As a result, NMP execution cannot proceed independently and must repeatedly exchange intermediates with the xPU across PCIe for residual and fusion computations. As discussed in Section~\ref{subsec:mot_challenges}, this back-and-forth communication leads to $2\times$ PCIe traffic amplification in the worst case, largely negating the performance gain of caching. Once cache overflow occurs, the key bottleneck is therefore no longer only redundant computation, but also the communication and serialization overhead introduced by the cache. 

To address this bottleneck, we propose \name, an algorithm-hardware co-designed \textbf{\underline{C}ompute-Cache \underline{O}perator} \textbf{\underline{D}is\underline{a}ggregation} architecture. \name keeps compute-intensive operators (compute operators) on the high-throughput xPU while offloading capacity-critical, bandwidth-bound {cache operators} to a specialized NMP subsystem, as reflected in Figure~\ref{fig:teaser}(c). On top of this physical disaggregation, \name uses a static-dynamic hybrid hardware-aware scheduler to reorganize fragmented cache activity into {coalesced segments}, reducing fine-grained compute-cache interleaving and cross-PCIe communication. It further exploits the independence between the two Classifier-Free Guidance (CFG) branches to overlap xPU compute with cache-side execution, thereby reducing exposed overhead and end-to-end latency. Through this cross-stack co-design, \name turns cross-timestep caching from nominal algorithmic reuse into real end-to-end system gains on edge platforms.

In summary, this paper makes the following contributions:
\begin{itemize}[leftmargin=*]
    \item \textbf{Hybrid Hardware-Aware Caching Scheduler:} We propose a scheduler that reorganizes fragmented cache activity into {coalesced segments}, reducing fine-grained compute-cache interleaving and cross-interface communication. We further introduce Dynamic Runtime Adjustment to safeguard rare volatile cases.
    \item \textbf{Lightweight DIMM-NMP Architecture:} We design a lightweight DIMM-side NMP subsystem that enables compute-cache operator disaggregation by executing memory-bound cache operators near memory with minimal hardware overhead.
    \item \textbf{CFG-Interleaved Pipelining:} We exploit CFG branch independence to overlap xPU compute with cache-side execution, thereby hiding exposed communication and NMP latency.
\end{itemize}

\noindent{Extensive experiments demonstrate that CODA achieves up to 1.80$\times$ end-to-end speedup and 1.74$\times$ higher energy efficiency, while incurring minor quality degradation over the SOTA baseline~\cite{zhao_pab_2025}.}

\section{Background}
\label{sec:background}

\subsection{Video Diffusion Models (VDMs)}\label{subsec:bg_vdm}
\begin{figure}[t]
    \centering
    \includegraphics[width=0.99\linewidth]{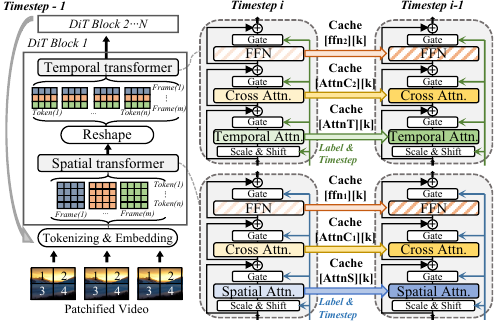}
    \caption{DiT blocks of VDMs (left). Cross-Timestep Caching, which replaces selected operators with a cache path (right).}
    \label{fig:DiT_and_Cache}
\end{figure}

As the generation fidelity of Video Diffusion Models (VDMs)~\cite{openai2025sora2,bytedance2025seedance,ali2025wan,kong2025hunyuanvideo} continues to improve, deploying them on edge devices for offline availability and privacy-preserving inference has become increasingly attractive~\cite{li_edge_2020,noauthor_mobilediffusion_nodate,zheng_diffusion_2025, kanipakam_privacy-preserving_2025}. However, this high fidelity relies on Diffusion Transformers (DiTs)~\cite{DiT2023,ma_latte_2024,zheng_open-sora_2024}. As illustrated in Figure~\ref{fig:DiT_and_Cache} (left), mainstream DiT architectures employ alternating Spatial and Temporal Transformers to process complex 3D spatial-temporal tokens. Because it is stateless across different iterative denoising steps, the DiT must execute extremely dense GEMM operators at every single step. This extreme compute-bound characteristic leads to very long generation latency at the edge~\cite{huang_vbench_2023,zheng_open-sora_2024,wan_wan_2025}.

\subsection{Cross-Timestep Caching (CTC)}\label{subsec:bg_caching}
To reduce redundant computation in iterative denoising, prior works have proposed Cross-Timestep Caching (CTC)~\cite{ToCa2025_localfix,selvaraju_fora_2024,zhao_pab_2025,fan_taocache_2025,liu_reusing_2025,wu_invardiff_2025,liu_speca_2025,cui2026bwcache,liu_freqca_2025,liu_fastcache_2026,ma_model_nodate2025}. As shown in Figure~\ref{fig:DiT_and_Cache} (right), CTC exploits feature-map similarity between adjacent denoising timesteps by caching the outputs of selected operators from the previous timestep and reusing them.
\revision{The ``operator'' refers to a module-level computation unit such as Attention or FFN. It includes primitive kernels such as LayerNorm, GELU, and Softmax inside the corresponding module, rather than treating these primitive kernels as separate cache operators.}
Importantly, cache reuse does not imply that an entire operator is completely skipped. Instead, it replaces the original compute-intensive GEMM operators with {cache operators} dominated by data retrieval and lightweight fusion. These operators typically fetch cache from storage, perform element-wise operations with lightweight features in the current branch (e.g., residual addition, scaling, or gated fusion), and then pass the updated features to subsequent serial operators. Therefore, while CTC reduces computation, it also shifts the corresponding execution toward a more memory-bandwidth-intensive pattern.

\begin{figure*}[t!]
    \centering
    \includegraphics[width=0.99\textwidth]{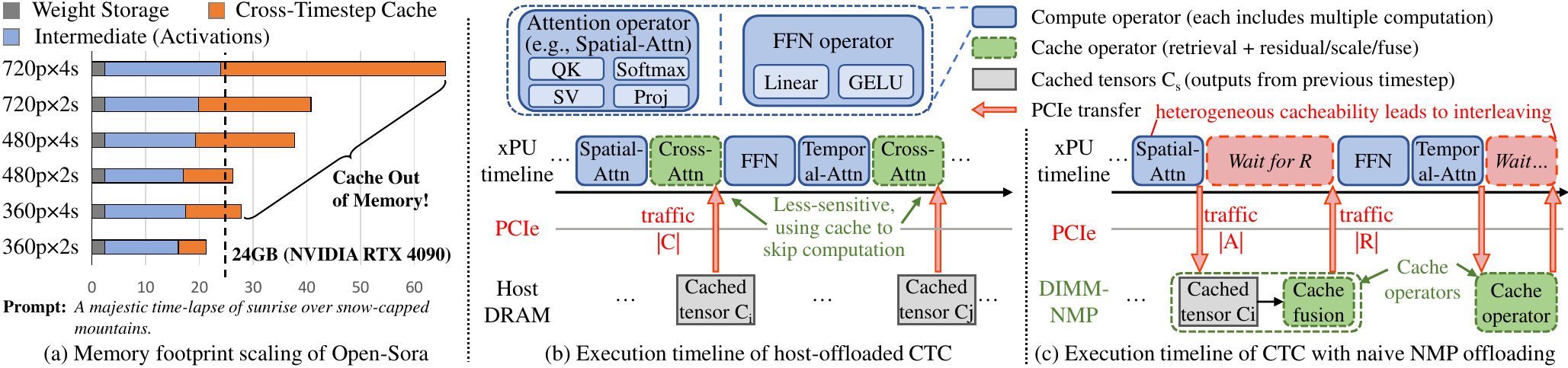}
    \caption{Motivation and insight. (a) Cache storage quickly exceeds edge-GPU VRAM as video size grows. \revision{(b) With host-offloaded CTC, the xPU fetches cached tensors from host DRAM and performs the cache-side residual/scaling/fusion locally. (c) With naive NMP offloading, the current activation must be sent to the DIMM-NMP and the fused result must return to the xPU before downstream compute operators can continue, causing repeated PCIe transfers and stop-and-wait serialization.}}
    \label{fig:Memory_Footprint}
\end{figure*}

\subsection{Classifier-Free Guidance (CFG)}\label{subsec:bg_cfg}
To improve fidelity and alignment between generated outputs and text prompts, modern VDMs widely adopt the Classifier-Free Guidance (CFG)~\cite{ho_classifier-free_2022, zheng_open-sora_2024,nava_meta-learning_2023,baykal_protodiffusion_2023,bradley_classifier-free_2024}. At each denoising timestep, the model executes two full branches in parallel: a conditional branch ($\epsilon_{cond}$) and an unconditional branch ($\epsilon_{uncond}$). At the end of that timestep, after both branches have completed the forward execution of all Transformer blocks, the system fuses their outputs using: $\epsilon = \epsilon_{uncond} + w \cdot (\epsilon_{cond} - \epsilon_{uncond})$
where $w$ is the guidance scale factor. The fused tensor serves as the timestep denoising result.

In existing implementations, these two branches are typically packed into a batch (size = 2) and submitted together to the GPU~\cite{chung2025cfg,saini2025rectified,phunyaphibarn_unconditional_nodate,li_towards_2025}. However, before reaching this fusion point, their computations within a timestep are data-independent from each other.

\subsection{Near-Memory Processing (NMP)}\label{subsec:bg_nmp}
Near-Memory Processing (NMP) is a promising architectural paradigm that reduces data movement by pushing computation closer to where data resides~\cite{asifuzzaman_survey_2023,park_attacc_2024,liu_enmc_2021,zhou_gnnear_2023,huangfu_medal_2019,gu_ipim_2020,li_ansmet_2025,LiSADIMM2025}. Existing NMP architectures are broadly divided by where compute logic is physically deployed.
{Intra-DRAM} designs (e.g., bank- or bank-group-level) integrate logic inside the DRAM die, close to sense amplifiers or bank I/O. Such designs exploit the extremely high internal bandwidth of DRAM but usually rely on customized DRAM processes and thus face stronger area and thermal constraints, as exemplified by UPMEM~\cite{gomez-luna_benchmarking_2022} and SK Hynix AiM~\cite{lee_aim_2022}.
{Extra-DRAM} designs (e.g., rank- or channel-level) instead place lightweight compute logic near the module buffer/register chip or memory controller without modifying the memory die. They are less intrusive to commodity memory, preserve the capacity advantage of commodity DRAM, and still provide high aggregated bandwidth. Representative DIMM-NMP systems include Samsung AxDIMM~\cite{Lee_AxDIMM_2022} and RecNMP~\cite{ke_recnmp_2020}, which accelerate large-capacity, low-arithmetic-intensity, memory-bound workloads such as data analytics~\cite{kim_darwin_2025} and LLM decoding~\cite{li_specpim_2024}.

\section{Motivation and Key Insight}
\label{sec:motivation}
\subsection{The Capacity and Bandwidth Crisis of CTC}
\label{subsec:mot_reversal}

Cross-Timestep Caching is fundamentally a space-for-time tradeoff, but its cache overhead is substantial. As analyzed in Figure~\ref{fig:Memory_Footprint}(a), the cache volume grows with video resolution and frame count. Even generating a 480p, 2-second video can already exceed the VRAM of a typical high-end edge GPU (e.g., 24~GB on an NVIDIA RTX 4090), making it difficult to keep the full cache on device~\cite{cache-dit_2025,videosys2024,yang_cogvideox_2025,kong2025hunyuanvideo}.

Existing systems therefore typically place overflow cache in host DRAM. This relieves the on-device capacity constraint but shifts the CTC bottleneck to data movement and serialized cache-path execution. \revision{Figure~\ref{fig:Memory_Footprint}(b) illustrates several serial operators within a cache-enabled timestep. CTC reuses cached outputs from the previous timestep and bypasses original computation, yet each cache operator still fetches the cache from host DRAM over PCIe and performs lightweight residual/scaling/fusion on the xPU before following operators proceed. Because these cache operators have very low arithmetic intensity, PCIe transfer latency can offset the benefit of skipping dense computation and may cause the host-offloading performance reversal shown in Figure~\ref{fig:teaser}(a).}

\subsection{\revision{Heterogeneous Operator Cacheability}}
\label{subsec:mot_heterogeneous_cache}

\revision{Our profiling results and prior CTC studies~\cite{ToCa2025_localfix,selvaraju_fora_2024,zhao_pab_2025} commonly observe that different operator types have different sensitivity to cross-timestep reuse. For example, Cross-Attention is often more stable across adjacent denoising timesteps, whereas FFN and some Spatial/Temporal Attention operators can be more sensitive to reuse error. As a result, CTC policies that focus on algorithm-level heterogeneous cacheability typically assign different caching aggressiveness to different operator types or network locations.}

\revision{Although such operator-aware policies help control generation-quality loss, they naturally create an interleaved execution pattern. Using the cache-enabled execution interval in Figure~\ref{fig:Memory_Footprint}(b)(c) as an example, Cross-Attentions are turned into cache operators, while neighboring operators remain on the native xPU compute path, forming an execution chain with interleaved dependencies between the compute path and the cache path. This interleaving is common in CTC execution, but it also exposes a limitation: cache operators are still sandwiched between dependent compute operators, dividing the execution flow into fine-grained communication boundaries.}

\subsection{The Pitfalls of Naive NMP Offloading}
\label{subsec:mot_challenges}

{Cache operators are typical bandwidth-bound operators: they frequently access large cached tensors, while their local computation mainly consists of lightweight element-wise operations. Therefore, a natural and necessary solution is to introduce a lightweight NMP substrate that provides an execution space for these memory-bound cache operators close to the cache storage. This near-memory execution substrate forms the hardware foundation for compute-cache operator disaggregation, enabling the cache path to be separated from the dense xPU compute path and providing the necessary condition for independent scheduling and overlapped execution.}

However, naive NMP offloading only changes where each cache operator is executed; it does not change the interleaved dependency structure described above. \revision{As shown in Figure~\ref{fig:Memory_Footprint}(c), an offloaded cache operator is still placed between adjacent xPU compute stages. Downstream compute can continue only after the memory-side fusion finishes and returns the result to the xPU. \textbf{Therefore, although NMP provides the execution substrate for cache-side execution, fine-grained serial dependencies still hinder compute-cache operator disaggregation.}}

This fine-grained coupling also amplifies cross-interface traffic. In host-offloaded CTC, the xPU mainly fetches the cached tensor $C$ over PCIe. Naive NMP instead makes each cache operator send the current activation $A$ to the DIMM-NMP and return the fused result $R$. Since $A$, $R$, and $C$ are typically comparable in size, this round trip can offset the benefit of near-memory execution. \revision{This motivates CODA to use a coarser cache-path granularity, amortizing one activation/result round trip across multiple cache operators instead of repeating it at every operator boundary.}

\revision{Another implication is exposed critical-path latency.} Naive NMP removes lightweight cache fusion from the xPU, but the xPU still waits at each cache-operator boundary before downstream dense computation can proceed. The added PCIe round trip and synchronization can outweigh the saved lightweight-operation time. \revision{CODA therefore needs to reorganize execution so that cache-side execution overlaps with dense xPU computation, rather than merely moving cache fusion closer to memory.}

\section{\name Design}
\label{sec:design}

\subsection{\name Overview and Execution Flow}
\label{subsec:overview}

\subsubsection{Cross-Stack Synergy}\ \par

\begin{figure}[t]
    \centering
    \includegraphics[width=0.99\linewidth]{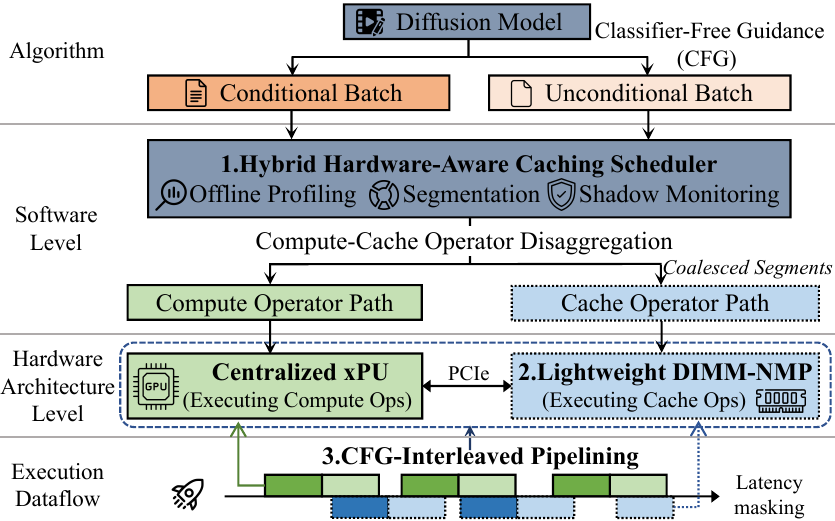}
    \caption{CODA overview and cross-stack synergy.}
    \label{fig:coda-overview}
\end{figure}

\erase{To address the communication and serialization bottlenecks introduced by cross-timestep caching in edge video generation, we propose \name. }As illustrated in Figure~\ref{fig:coda-overview}, \name is organized around compute-cache operator disaggregation. It begins with the \textbf{Hybrid Hardware-Aware Caching Scheduler}, which makes the originally intertwined execution flow explicit by separating it into a compute path and a cache path, while reorganizing the latter into coarse-grained coalesced cache segments to amortize communication overhead. This separation enables a clean hardware mapping, in which compute-intensive dense operators remain on the centralized xPU, while memory-bound cache operators are assigned to the \textbf{Lightweight DIMM-NMP}. Importantly, this near-memory subsystem is intentionally streamlined for the low-arithmetic-intensity access and fusion patterns of cache operators, providing a low-overhead execution substrate for the disaggregated cache path. Building on this separation and mapping, \textbf{CFG-Interleaved Pipelining} exploits the independence of the two CFG branches before final fusion, allowing the cache-side DMA/NMP execution of one branch to overlap with the xPU dense computation of the other. In this way, the remaining communication and processing costs of the cache path can be largely hidden behind the xPU compute window. \erase{These three components therefore form a closed system pipeline rather than isolated optimizations: the scheduler exposes and restructures the cache path, the lightweight DIMM-NMP physically decouples its execution from the dense compute path, and the CFG-interleaved pipeline converts the residual cache-path overhead into latency that can be masked. The following subsections describe the corresponding scheduling, architecture, and execution.}

\begin{figure}[t]
    \centering
    \includegraphics[width=0.99\linewidth]{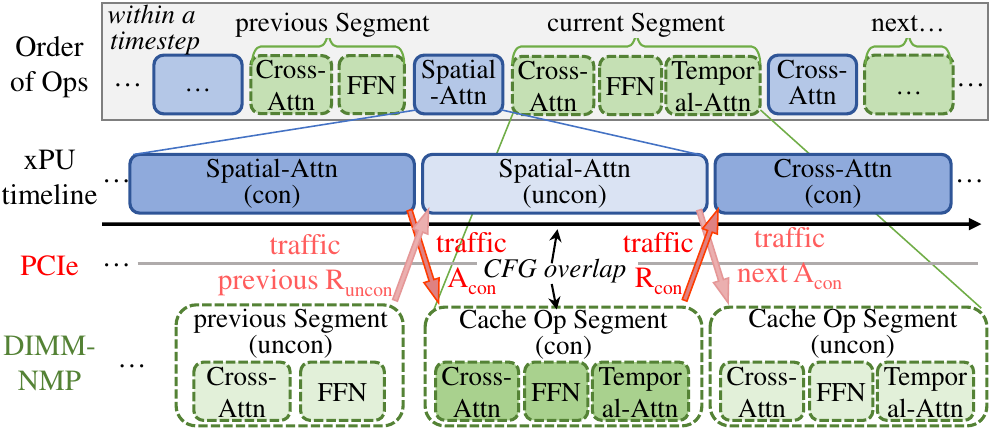}
    \caption{\revision{\name execution walkthrough. Coalesced segments are formed within a timestep, execute cache-side operators on DIMM-NMP, and overlap with dense xPU computation on the other CFG branch before timestep-level CFG fusion.}}
    \label{fig:coda-walkthrough}
\end{figure}
\subsubsection{\revision{End-to-End Execution Walkthrough}}\ \par
\label{subsubsec:coda_execution_walkthrough}

\revision{Figure~\ref{fig:coda-walkthrough} illustrates the execution flow of \name within a cache-enabled timestep. The top row shows the operator order in the current timestep. CODA scheduler combines operator cacheability with hardware cost to adjust the cache policy and form coalesced cache segments for cache-side execution. These segments read cached tensors produced by previous timesteps, but the segment does not cross timestep boundaries, because CFG must synchronize the two branches and perform fusion at the end of each timestep.}

\revision{When execution enters a coalesced cache segment, the xPU no longer offloads each cache operator separately. Instead, it sends the current activation to the DIMM-NMP through PCIe at the segment boundary. The DIMM-NMP then reads the corresponding cached tensors from memory and executes the segment's cache retrieval, residual/scale/fusion operations on the memory side. After the segment finishes, the fused result is returned to the xPU and the dependent downstream compute operators continue. In this way, activation/result transfers are consolidated at segment boundaries, amortizing PCIe communication and synchronization overhead.}

\revision{\name further hides cache-side latency through CFG execution. Within a denoising timestep, the conditional and unconditional branches are independent before final CFG fusion. Thus, while one branch's coalesced cache segment runs on the NMP, the xPU can advance dense computation on the other branch. The branches synchronize only after all operators in the timestep finish, when CFG fusion produces the final denoising output. This walkthrough sets the semantics used below: the scheduler forms coalesced segments, NMP executes them near memory, and CFG-interleaved pipelining overlaps cache-side execution with xPU dense computation.}

\subsection{Hybrid Hardware-Aware Caching Scheduler}
\label{sec:hybrid_scheduler}

\begin{figure}[t]
    \centering
    \includegraphics[width=0.99\linewidth]{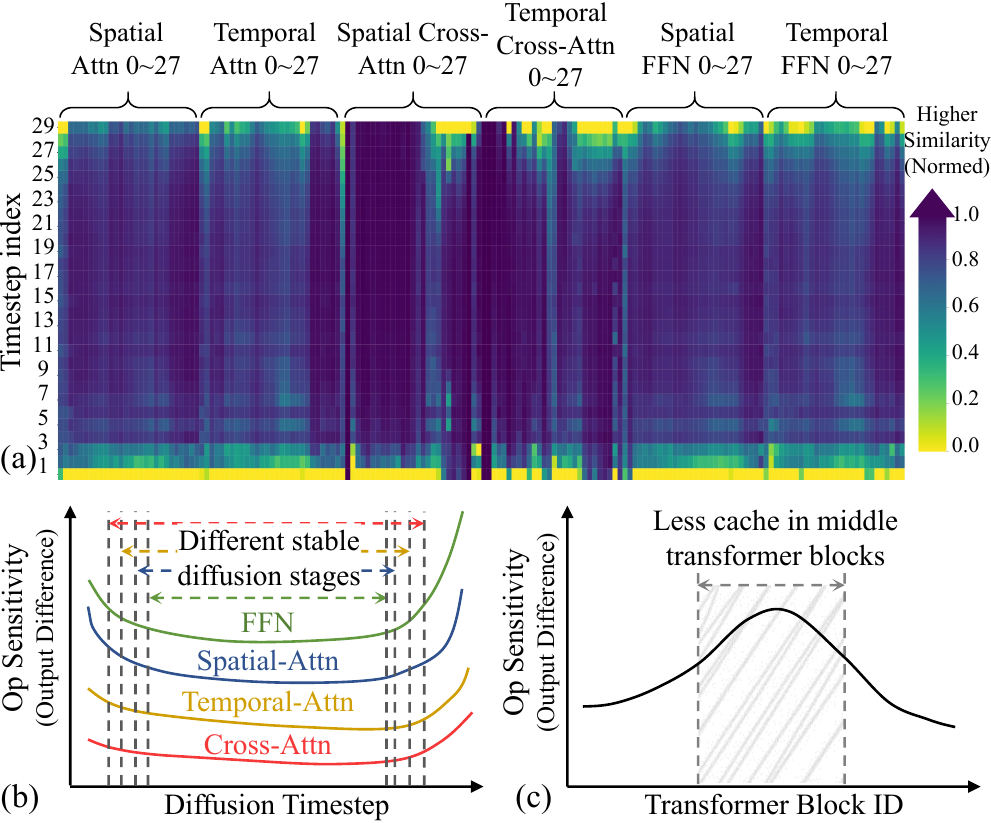}
    \caption{Spatial-temporal heterogeneity of cacheability: (a) Cross-timestep output stability (L1 similarity) profiling of different operators in Open-Sora~\cite{zheng_open-sora_2024}  (Normalized across all timesteps for each op). (b) Relative operator sensitivity across diffusion timesteps: FFN$>$Spatial$>$Temporal$>$Cross. (c) Operators in the middle blocks are generally more sensitive than operators of the same type near either end of the network.} 
    
    \label{fig:heatmap}
\end{figure}

\phantomsection\label{rev:ctc_policy_novelty}
\revision{\name's hardware-aware caching scheduler extends prior CTC cache policies from reuse selection to hardware-aware execution planning. Prior CTC methods~\cite{zhao_pab_2025,selvaraju_fora_2024} choose reusable timesteps and operators for quality-preserving computation reduction. \name uses the operator-instance similarity profile, which captures the above operator/timestep differences as well as the backbone-depth variation revealed by our profiling in Figure~\ref{fig:heatmap}(c), and weighs it together with xPU-DIMM communication and DIMM-NMP execution/overlap costs. The scheduler therefore determines cached operator instances and the resulting coalesced segment boundaries offline, while runtime only applies lightweight corrections when an input deviates from the offline profile.}

\subsubsection{Static Profiling-Based Scheduling Policy}\label{subsubsec:scheduler_static_policy}\ \par
\label{rev:scheduler_candidate_details}
\revision{Given the profiled operator sequence of each model, \name searches cache decisions offline, treating each operator instance as a candidate. Consecutive selected candidates within the same timestep form coalesced cache segments, and the remaining xPU operators separate these segments to define xPU-DIMM communication boundaries. This reduces fine-grained PCIe transactions while controlling generation-quality loss.}

\noindent\textbf{\revision{Cacheability quantification for scheduling.}}
\phantomsection\label{rev:cacheability_quantification}
\revision{For static policy, \name profiles each model with 10 representative prompts before inference and measures each operator instance's normalized output L1 similarity across adjacent denoising timesteps. The averaged profile is used as the offline cacheability signal because cacheability mainly reflects model-structural factors such as operator type, timestep, and network depth and is relatively stable across inputs. For rare outlier inputs, Dynamic Runtime Adjustment checks runtime feature evolution and applies a more conservative cache policy.}

\phantomsection\label{rev:coalesced_segmentation_details}
\noindent\textbf{Coalesced segmentation.}
Based on the cacheability signal above, the scheduler searches for a static cache policy $S$. A coalesced segment is a continuous sequence of cache operators \revision{within the same timestep} that is packed and offloaded to the NMP as a single transaction \revision{and executed in the original dependency order}. Its key benefit is to reduce fine-grained interleaving between compute operators and cache operators: instead of incurring PCIe communication for every cached operator, the system communicates mainly at the input and output boundaries of each segment.
To jointly balance reuse benefit and execution overhead, the scheduler determines this static policy $S$ by minimizing the following objective:
\begin{equation}
E(S) = w_{cost} \cdot \sum \text{Cost}(S) + w_{frag} \cdot \text{Fragment}(S)
\end{equation}
where $\text{Cost}(S)$ captures \revision{the fidelity risk of the cached operator instances selected by $S$}. For each cached operator, we use the profiled average cross-timestep output instability (implemented as $1-\mathrm{Similarity}_{L1}$) as a proxy
for the error introduced by replacing native computation with caching, and accumulate it over the operators covered by the plan. In contrast, $\text{Fragment}(S)$ penalizes both overly fragmented segments and segments whose execution latency deviates from the hardware-preferred pipeline window $T_{optimal}$ (which is further explained in Section~\ref{subsec:CFG-pipe}):
\begin{equation}
\text{Fragment}(S) = \sum_{i=1}^{k} \left( C_{comm} + w_{dev}\cdot \max(0, T_{exec}(S_i) - T_{optimal}) \right)
\end{equation}

\begin{figure*}[t]
    \centering
    \includegraphics[width=0.99\linewidth]{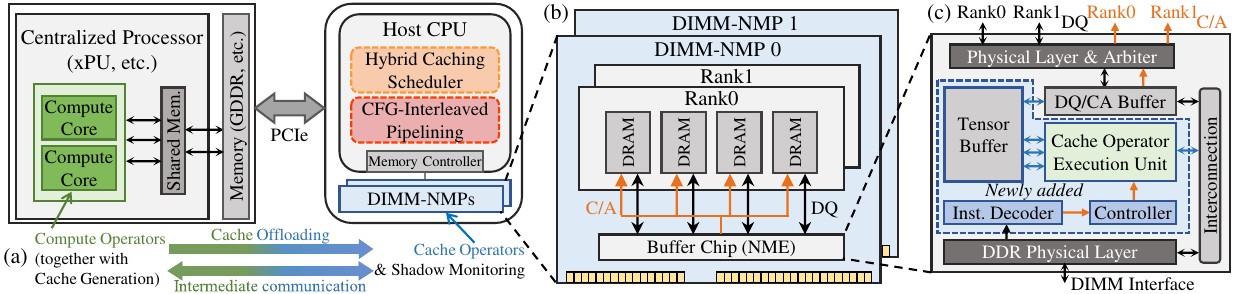}
    \caption{\name architecture: (a) Compute-cache operator disaggregation between the xPU and DIMM-NMP subsystem. (b) Rank-level, buffer-chip-based DIMM-NMP organization. (c) Near-Memory Engine (NME) for cache-based diffusion operators.}
    \label{fig:System_overview}
\end{figure*}
$C_{comm}$ denotes the base communication overhead of a single PCIe transaction, which naturally discourages excessively fragmented segments, and $T_{exec}(S_i)$ is the estimated execution latency of segment $S_i$ on the NMP. The objective therefore jointly constrains two risks: overly aggressive caching that may harm generation quality, and overly fragmented or pipeline-misaligned segments that undermine system efficiency. \revision{These weights are tunable across models and policy preferences. By default, \name uses $w_{cost}=3$, $w_{frag}=0.8$, and $w_{dev}=1$, where $w_{cost}$ favors lower reuse-risk operators, $w_{frag}$ favors fewer and longer segments, and $w_{dev}$ penalizes segments exceeding the CFG-overlap window.} \revision{Since the search space over cache decisions and induced segment boundaries} is non-convex, we use simulated annealing~\cite{SA} to optimize the cache decisions and generate an offline static schedule for each model. \revision{In our implementation, the target cache ratio controlling overall caching aggressiveness is set to 25\%\textasciitilde 50\% by default depending on the model. Simulated annealing uses 300K\textasciitilde 500K proposals per model, an initial temperature of 1.0, a cooling factor of 0.995 every 1K proposals, and early stopping after 20K proposals without objective improvement. Because the search only uses profiled statistics and analytical latency estimates, it finishes in minutes as a one-time offline cost and does not affect the inference critical path.}
\subsubsection{Dynamic Runtime Adjustment}\label{subsubsec:dynamic_runtime_adjustment}\ \par
Although the static policy covers the vast majority of inputs, real-world inference may still encounter a small number of prompts whose early-step feature evolution deviates noticeably from the offline profiling pattern, making the planned caching start point overly aggressive. To handle such cases, \name introduces an asynchronous runtime safeguard on top of the static schedule to validate the planned cache-entry point before entering the caching phase.

\noindent\textbf{Shadow monitoring.}
This mechanism exploits a natural system window. During the high-noise early denoising stage, feature variation is typically too large for caching to be safely enabled, so the NMP remains idle while the xPU continues along the native execution path. We use this window to asynchronously offload the output features of timesteps $t+2$ and $t+1$ to the NMP before the scheduled caching start step $t$, and compute their L1 distance in the background. Because this check is decoupled from the main execution path, it introduces no exposed latency on the critical path and serves as a lightweight shadow monitor.

\noindent\textbf{Fixed-step fallback.}
If the measured distance exceeds a predefined safety threshold, the runtime interprets it as a signal that the operator remains too unstable near the current cache-entry point to safely enter the caching phase. In that case, the host runtime temporarily disables caching for that operator and falls back to native xPU execution for \revision{$N_{\mathrm{fb}}$} nearby denoising timesteps (typically $t$ and $t-1$). We adopt this fixed-step fallback instead of continuous online monitoring because it is sufficient to cover the common instability cases around cache entry while avoiding the additional control complexity, storage-access interference, and hardware overhead of continuous monitoring. In this way, DRA preserves the efficiency of the static schedule in the common case while providing a low-overhead runtime correction for rare volatile inputs.

\subsection{\name Architecture and NMP Subsystem}
\label{sec:coda_architecture}

To efficiently execute the coarse-grained cache plans generated by the scheduler in the previous section, \name requires a hardware substrate that supports both operator disaggregation and streaming data organization. This section presents the overall system architecture of \name, the NMP microarchitecture tailored to cache operators, and the execution dataflow that remains compatible with modern memory protocols.

\subsubsection{System Architecture and Design Rationale}\label{subsubsec:architecture_rationale}\ \par
As shown in Figure~\ref{fig:System_overview}(a), \name adopts a heterogeneous execution framework based on compute-cache disaggregation. The xPU serves as the centralized compute engine and executes the compute-intensive dense operators in the DiT backbone, while the memory-bound cache operators are mapped to the DIMM-NMP subsystem, where they are executed on the memory side and interact with the xPU only at segment boundaries.

CODA adopts the rank-level, center-buffer-based extra-DRAM NMP shown in Figure~\ref{fig:System_overview}(b)(c) as a balanced design point for cache-path disaggregation. \revision{This choice reflects the cost-benefit tradeoff: providing an independent cache-side execution space without introducing complex control or intrusive DRAM-die modifications, while keeping its latency largely hidden by dense xPU computation.}

First, the rank-level organization better matches CODA's tiled multi-stream fusion flow with lower integration complexity. For each tile group, the NME coordinates the incoming activation stream with multiple aligned cache streams $(C_1,\dots,C_n)$ and applies the fused operations in order. Placing this coordination at the rank-level center buffer avoids introducing fine-grained bank-local controllers, strict stream co-location constraints, or cross-bank partial movement and synchronization. \revision{It also preserves a commodity-DIMM-like organization without modifying DRAM dies, reducing area, thermal, verification, and integration costs.}

\revision{Second, rank-level NMP is sufficient for CODA's cache-path execution. For a coalesced segment, increasing internal DRAM bandwidth mainly shortens tiled fusion in the NME, while segment-boundary activation/result transfers and synchronization remain. In our evaluated configuration, tiled fusion is less than 20\% of estimated cache-path latency for common segment lengths $n=2/3/4$, so Amdahl's Law limits the gain from replacing the rank-level NME with near-bank NMP. Our evaluation (Section~\ref{subsec:ablation}) further shows that the remaining PCIe/NMP overhead can be largely hidden behind xPU computation. Thus, near-bank PIM is an optional more aggressive design point, with limited extra end-to-end benefit relative to its hardware and control complexity.}

\subsubsection{NMP Microarchitecture}\label{subsubsec:nmp_microarchitecture}\ \par
Based on the above architectural choice, CODA further streamlines the NME microarchitecture according to the characteristics of cache operators. As shown in Figure~\ref{fig:EU_Dataflow}(b), these operators are fundamentally dominated by element-wise scaling and fusion, and we therefore adopt a SIMD Processing Element (PE) tailored to this execution pattern.
\revision{In Figure~\ref{fig:EU_Dataflow}(b), $m$ denotes the number of PEs in one NME, and we set $m=4$ according to the DSE in Section~\ref{subsec:architecture_dse}. $n$ denotes the number of cached tensors/operators fused within the current coalesced segment.}
Data entering the PE does not always follow a unified scale-and-fuse path. For gated fusion, the input first passes through a scalar multiplier, where it is multiplied by a broadcast scaling factor, and is then fused through the floating-point adder. For the more common residual-only case, all scaling multipliers are uniformly bypassed, and the input data are directly forwarded to the subsequent accumulation stage. Meanwhile, operand isolation clamps the multiplier inputs to stable constants to suppress redundant switching and reduce dynamic power.

\phantomsection\label{rev:dra_host_cpu_reduction}
In addition to executing cache operators, this microarchitecture must also support the shadow monitoring mechanism introduced in the previous section. To this end, CODA reuses the existing datapath instead of introducing dedicated monitoring hardware. \revision{For DRA, the NME streams the two feature tiles to be compared through the PE lanes and bypasses the multiplier path. The augmented floating-point adder computes the element-wise difference and absolute value through sign-bit manipulation. The firmware then reorganizes the resulting difference vector and feeds it back to the same PE datapath, where the adder and registers perform tree-style accumulation to produce a local L1-difference scalar within each DIMM.} Since the local computations are independent, the final global reduction can be directly handled by the host CPU when reading out the local scalars from all DIMMs. \revision{This scalar-level reduction scales only with DIMM count and is negligible compared with cache tensor movement. The reused datapath computation and final scalar aggregation run in the cache-entry shadow window and can be hidden by the xPU's dense early-stage computation, adding no exposed critical-path latency. This monitoring occurs before cache is enabled, when early denoising features are unstable and unsuitable for cache reuse, so it does not run in parallel with cache operators and requires no separate NMP partition.}

\subsubsection{Firmware Execution: Tiled Operator-Fusion Dataflow}\label{subsubsec:tiled_operator_fusion}\ \par
Although cache operators process large amounts of data, their execution pattern remains dominated by element-wise scaling and fusion rather than high-arithmetic-intensity matrix operations. This operator structure allows CODA to partition them consistently by spatial position and execute them in a streaming manner across DIMM-NMPs. At the same time, a single cache operator can still involve hundreds of megabytes of data, while the on-chip SRAM capacity inside the NME remains highly limited. CODA therefore adopts a tiled operator-fusion dataflow.

\begin{figure}[t]
    \centering
    \includegraphics[width=0.995\linewidth]{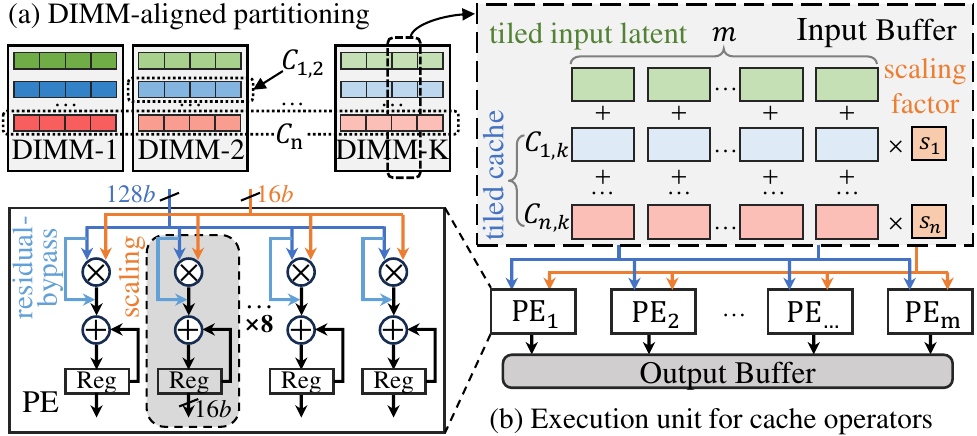}
    \caption{(a) DIMM-aligned partitioning of the input activation and cached tensors within a coalesced segment for parallel execution across DIMM-NMPs. (b) Streamlined execution unit and PE datapath for cache-based diffusion operators. \revision{(n is the coalesced segment length, and m is the number of PEs.)}}
    \label{fig:EU_Dataflow}
\end{figure}

As shown in Figure~\ref{fig:EU_Dataflow}(a), within a coalesced segment, the runtime first performs DIMM-aligned partitioning along the spatial dimension for both the input activation and the cached tensors \revision{from the latest reusable timestep}, so that they are split at the same positions and mapped to different NMPs. Each DIMM therefore processes only its assigned input partition together with the aligned cache partitions \revision{for the $n$ cache operators in the current coalesced segment}. \revision{Within each DIMM, these local partitions are further divided into aligned tile groups. Each tile group contains multiple contiguous SIMD-width chunks, where each chunk matches the PE-array width, i.e., $m$ PEs $\times$ 8 lanes. The number of chunks in a tile group is selected according to the fusion depth $n$ and the available NME buffer capacity (64 chunks in our evaluated configuration).} During execution, the xPU first writes the partitioned input activation into DRAM through standard write requests, \revision{and the NME firmware then iterates over these aligned tile groups.}

\revision{Within a tile group, the firmware processes data chunk by chunk. For each SIMD-width chunk, the PE array loads the current activation chunk into its lanes. It then streams the corresponding chunks from the aligned cache tiles in the original segment order. After each scaling, residual, or fusion update, the partial vector remains in the PE registers and is directly consumed by the next fused operation. Only after all $n$ cache operations for this chunk are completed is the final output chunk written to the Output Buffer. The firmware then advances to the next chunk in the tile group, and finally to the next tile group. After all tile groups are processed, the Output Buffer contains the final fused segment output to be returned to the xPU. CODA therefore avoids materializing full-tensor or full-tile intermediate results after every cache operator, reducing internal bandwidth overhead and unnecessary write-backs within the NME.}

\phantomsection\label{rev:conflict_aware_layout}
The above tiled dataflow triggers concurrent accesses to multiple cache tiles within the inner loop. To reduce potential row/bank conflicts caused by highly regular contiguous layouts under concurrent accesses, CODA further applies conflict-aware placement during tensor layout to decorrelate the low-level address mapping of cache tensors that are accessed together. \revision{Concretely, instead of changing the logical tensor order, CODA assigns the $n$ DIMM-local tiled cache tensors consumed by the same coalesced segment to different base regions or allocation offsets when possible. This reduces the likelihood that their low-level bank/bank-group mappings align repeatedly under the memory-controller address mapping, while preserving the element-wise tiled execution order.} This treatment is only an engineering layout strategy to support the tiled operator-fusion dataflow more robustly, rather than an independent method.

\subsection{CFG-Interleaved Pipelining}
\label{subsec:CFG-pipe}

\begin{figure}[t]
    \centering
    \includegraphics[width=0.99\linewidth]{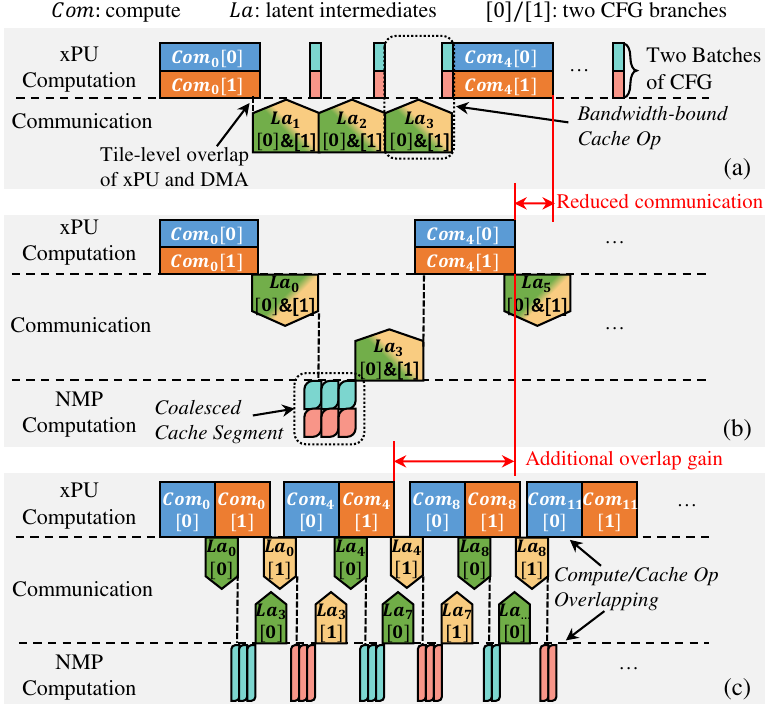}
    \caption{\name execution evolution: (a) Scheduler-only flow, where coalesced cache segments still stall the xPU through bandwidth-bound transfers. (b) Adding NMP offloading reduces the communication span of the coalesced cache segment, but the xPU remains idle during cache-path execution. (c) CFG-interleaved pipelining overlaps one branch with the other branch's coalesced cache segment running on the NMP.}
    \label{fig:dataflow_evolution}
\end{figure}

Figure~\ref{fig:dataflow_evolution} illustrates the execution evolution enabled by \name. Figure~\ref{fig:dataflow_evolution}(a) shows the scheduler-only flow: after the Hybrid Hardware-Aware Caching Scheduler reorganizes fragmented cache operators into hardware-friendly coalesced segments, the cache path still executes through bandwidth-bound transfers and stalls the xPU. Figure~\ref{fig:dataflow_evolution}(b) then maps these coalesced cache segments to the NMP. Compared with the scheduler-only flow in Figure~\ref{fig:dataflow_evolution}(a), this naive ``offload-and-wait'' execution shortens the exposed communication span of cache processing, but the cache path and xPU compute still proceed in a serialized manner, leaving system-level pipeline bubbles. To remove this serialization bottleneck, \name introduces CFG-Interleaved Pipelining, as shown in Figure~\ref{fig:dataflow_evolution}(c).

As discussed in Section~\ref{subsec:bg_cfg}, the conditional and unconditional branches of the CFG remain independent before the final fusion within each denoising timestep. \name exploits this structure by breaking the bundled dispatch paradigm (Batch=2) used in existing frameworks and unrolling it into two asynchronous execution streams at the system level. This split introduces a small kernel launch overhead and slightly reduces the computational density of each dispatch. Figure~\ref{fig:dataflow_evolution}(c) reflects this effect: the computation block of a single branch is slightly longer than exactly half of the two-branch consolidated block in Figure~\ref{fig:dataflow_evolution}(a). Nevertheless, this overhead is amortized by the resulting cross-branch overlap. This also gives the runtime enough separation to turn branch independence into a practical masking window to hide cache-side latency.

In execution, the xPU first starts one CFG branch (the blue blocks in Figure~\ref{fig:dataflow_evolution}(c)). To avoid waiting for an entire monolithic operator to finish before communication begins, \name applies explicit macro-tiling to GEMM operators. \revision{For GPU instances, this macro-tiling exposes runtime-visible subGEMM boundaries. Once a subGEMM materializes its output tile in GPU memory, the runtime uses multi-stream asynchronous dispatch and event synchronization to launch the corresponding DMA without waiting for the entire GEMM to finish.} This creates a tile-level pipeline between xPU execution and data movement. Meanwhile, after receiving the first data tiles, the NMP can immediately start its internal tiled operator-fusion dataflow, while the xPU proceeds to the other CFG branch (the orange blocks) instead of stalling on the cache path.
    
These activities can overlap in time: downstream DMA for the current operator (xPU $\rightarrow$ NMP), upstream DMA for the preceding cache result (NMP $\rightarrow$ xPU), dense xPU computation, and cache-side NMP execution. More importantly, Figure~\ref{fig:dataflow_evolution}(c) shows from $Com_0$ to $Com_4$ that the NMP processes a continuous coalesced segment generated by the hybrid scheduler, rather than scattered cache operators triggered one by one. Accordingly, this overlap is built directly on the coarse-grained cache reorganization in Section~\ref{sec:hybrid_scheduler}. Ideally, by the time the xPU finishes the other CFG branch, the NMP has also completed the corresponding coalesced segment, enabling a tight handoff between the two branches. Accordingly, the orange branch provides the xPU compute window that masks the cache segment opened by the preceding blue branch. Under this view, the target $T_{optimal}$ in the cost function of Section~\ref{sec:hybrid_scheduler} represents the desired NMP masking window for this overlap. Specifically, the scheduler aims to keep the end-to-end NMP latency---including both bidirectional PCIe communication and internal NMP computation---within the xPU branch compute window as much as possible. When this condition is satisfied, the exposed serial waiting time in Figure~\ref{fig:dataflow_evolution}(b) is substantially reduced, leading to higher xPU utilization.

\section{Evaluation}
\label{sec:evaluation}

\subsection{Experimental Setup}
\label{subsec:exp_setup}

\subsubsection{\name System}\label{subsubsec:evaluation_system}\ \par
We evaluate \name in a hardware-software co-designed environment. The centralized processor (xPU) is a single NVIDIA GeForce RTX 4090 with 24 GB GDDR6X memory. The host system uses an Intel Core i9-13900K and 128 GB DRAM. The GPU and the near-memory subsystem communicate over PCIe 4.0 x16, with an aggregate bidirectional bandwidth of approximately 64 GB/s.

For evaluation, GPU end-to-end latency and kernel latency are profiled using NVIDIA Nsight Compute, while cycle-level NMP performance is modeled by extending Ramulator 2.0 \cite{ramulator2}. \revision{We compute bidirectional PCIe transfer timing from the segment-boundary activation/result size and PCIe bandwidth. We then compose GPU, PCIe-transfer, and NMP events into a dependency timeline: dependent events are serialized, independent events may overlap, and final latency is taken from the critical path, which also determines hidden and exposed overhead.} We implement the NME microarchitecture in RTL and use Synopsys Design Compiler with the TSMC 7nm technology library to estimate area and power. The 64KB on-chip SRAM buffer is modeled using a commercial SRAM compiler. The evaluated DIMM-NMP configuration is summarized in Table~\ref{tab:ndp_config}.

\begin{figure*}[t]
    \centering
    \includegraphics[width=0.99\linewidth]{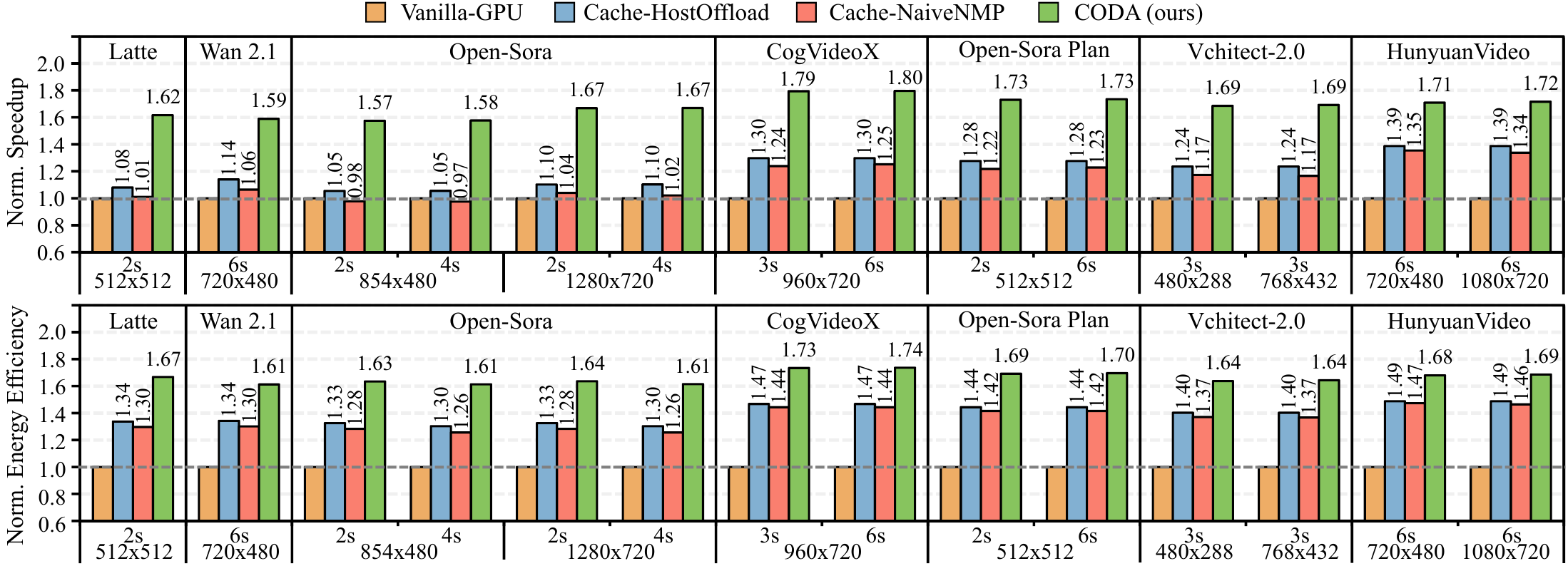}
    \caption{\name speedup and energy efficiency improvement on different models normalized to Vanilla-GPU.}
    \label{fig:speedup}
\end{figure*}

We estimate total energy as the sum of GPU, NME, memory-subsystem, and PCIe-link energy. NME dynamic and static power are obtained from RTL synthesis and the SRAM compiler, while memory and PCIe energy are estimated from memory-access statistics and link traffic. All latency, energy, and quality results are averaged over five runs for each prompt, and the reported end-to-end results include runtime overheads such as DRA handling.

\subsubsection{Baseline Systems}\label{subsubsec:baseline_systems}\ \par
\phantomsection\label{rev:cache_naivenmp_baseline}
We consider four system configurations. Vanilla-GPU denotes the original execution without caching. Cache-HostOffload offloads overflow cache to host DRAM when VRAM capacity is insufficient. Cache-NaiveNMP offloads cache-related operators to the near-memory side \revision{using the same DIMM-NMP hardware configuration as CODA}, but does not use our hardware-aware scheduler or CFG-interleaved pipelining. \revision{When the cache policy naturally selects adjacent cache operators, Cache-NaiveNMP dispatches these contiguous cache operators as a segment and communicates only at the input and output boundaries of that segment.} CODA denotes the full system, including the lightweight DIMM-NMP, the hardware-aware caching scheduler, and CFG-interleaved pipelining.

\phantomsection\label{rev:pab_baseline_mechanism}
We use PAB~\cite{zhao_pab_2025} as the CTC algorithmic baseline. \revision{PAB derives its cache policy from per-model profiling of cross-timestep similarity and applies it as fixed reuse rules over selected timesteps and operator types.} For system performance comparisons, Cache-HostOffload and Cache-NaiveNMP adopt the default PAB caching setup \revision{(i.e., PAB-Cons in Table~\ref{tab:eval_summary})}, while CODA uses its own caching policy. We therefore compare their end-to-end system gains at a similar video-quality operating point, rather than under a matched cache ratio or identical cached-operator coverage. For quality evaluation, we further report both the conservative and aggressive PAB settings to contextualize the quality-performance trade-off.

\subsubsection{Workloads and Metrics}\label{subsubsec:workloads_metrics}\ \par
We evaluate \name on a diverse set of VDMs, including Latte~\cite{ma_latte_2024}, Open-Sora~\cite{zheng_open-sora_2024}, Open-Sora Plan~\cite{opensora-plan}, Wan 2.1~\cite{wan_wan_2025}, HunyuanVideo~\cite{kong2025hunyuanvideo}, Vchitect-2.0~\cite{fan_vchitect-20_2025}, and CogVideoX-5B~\cite{yang_cogvideox_2025}. For models that support multiple resolutions and video lengths, we select several representative configurations. All text prompts are sampled from VBench~\cite{huang_vbench_2023} to cover diverse generation scenarios.

We report both system efficiency and generation quality. System efficiency is measured by end-to-end latency and total energy consumption. Quality is measured by VBench~\cite{huang_vbench_2023} \revision{for multi-dimensional video quality}, PSNR and SSIM \revision{for frame-level reconstruction fidelity}, and LPIPS \revision{for perceptual feature distance.} Each workload uses 100 prompts, and to reduce the impact of diffusion sampling stochasticity, all quality, latency, and energy results are averaged over five independent runs for each prompt.

\begin{table}[t]
\centering
\caption{Configuration Details of the Evaluated DIMM-NMP}
\label{tab:ndp_config}
\resizebox{\columnwidth}{!}{%
\begin{tabular}{c}
\hline
\textbf{NME Architecture} \\ \hline
Configuration: 4 PEs, 8 Lanes/PE, NME Buffer size: 64 KB, One NME per DIMM  \\
Area overhead: 0.029 $mm^2$ \textbar{} Power overhead: 10.17 mW \textbar{} @ 1 GHz  \\ \hline
\textbf{DIMM Parameters} \\ \hline
DDR4-3200, 16 GB/DIMM $\times$ 8, 2 DIMMs/channel \\
2 ranks/DIMM, 2 bank groups/rank, 4 banks/bank group \\ \hline
\textbf{DIMM Timing Specifications} \\ \hline
tRC=74, tRCD=22, tCL=22, tRP=22, tBL=4 \\
tCCD\_S=4, tCCD\_L=8, tRRD\_S=4, tRRD\_L=6, tFAW=26 \\ \hline
\end{tabular}%
}
\end{table}

\subsection{End-to-End Speedup and Energy Efficiency}
\label{subsec:e2e_performance}

Figure~\ref{fig:speedup} reports the normalized speedup and normalized energy efficiency under different model, resolution, and video-length settings. Overall, CODA achieves the highest performance and energy-efficiency gains across all evaluated configurations, showing that it can translate the potential computation savings of cache reuse into realized end-to-end system benefits.

A closer baseline comparison shows that moving cache operators to the memory side alone is insufficient. Cache-HostOffload reduces redundant GPU computation through cache reuse and improves over Vanilla-GPU, but only modestly. Cache-NaiveNMP underperforms Cache-HostOffload across all workloads and even falls below Vanilla-GPU for Open-Sora (854$\times$480), revealing a performance reversal. This shows that algorithmic reuse does not translate into speedup when the cache path remains tightly interleaved and serially dependent on the native compute path, leaving communication and serialization overhead on the critical path. CODA makes reuse beneficial by reorganizing the cache path into hardware-friendly execution and overlapping it with GPU computation.

Figure~\ref{fig:speedup} also shows that CODA's gain generally grows under heavier workloads, especially at higher resolutions. This follows from the scaling gap between attention computation and cached states. Let $T$ be tokens per frame, $F$ frames, and $d$ hidden dimension. Spatial attention scales as $F T^2 d$, temporal attention as $T F^2 d$, while the cache footprint grows roughly linearly with $F T d$. As resolution increases, cache reuse removes proportionally more GPU computation, explaining why gain amplification is more evident at higher resolutions than from merely increasing video length.

From the energy perspective, Cache-HostOffload and Cache-NaiveNMP often show larger energy-efficiency gains than speedup gains. Although they do not effectively hide waiting time from communication or near-memory execution, cache reuse still removes substantial redundant GPU computation. Because GPU computation dominates power, while PCIe transfers and lightweight NMP execution add smaller overhead, these baselines can save energy even with limited latency gains. CODA further reduces redundant GPU computation and hides exposed cache-step stalls through schedule reorganization and cross-branch pipelining, delivering the best end-to-end speedup and energy efficiency.

\begin{table}[t]
\centering
\setlength{\tabcolsep}{3pt}
\resizebox{\linewidth}{!}{
\begin{tabular}{cc|cccc}
\toprule
Model & Method & VBench $\uparrow$ & PSNR $\uparrow$ & SSIM $\uparrow$ & LPIPS $\downarrow$ \\
\midrule
\multirow{4}{*}{\shortstack{Latte \cite{ma_latte_2024}}} & Original & 0.8115 & - & - & - \\
 & PAB-Cons & 0.7873 & 20.3445 & 0.7101 & 0.2711 \\
 & PAB-Aggr & 0.7556 & 17.8648 & 0.6499 & 0.3905 \\
 & \cellcolor{gray!20}CODA & \cellcolor{gray!20}0.7916 & \cellcolor{gray!20}18.7831 & \cellcolor{gray!20}0.6688 & \cellcolor{gray!20}0.3237 \\
\midrule
\multirow{4}{*}{\shortstack{Open-Sora \cite{zheng_open-sora_2024}}} & Original & 0.8051 & - & - & - \\
 & PAB-Cons & 0.7783 & 27.6532 & 0.8891 & 0.0892 \\
 & PAB-Aggr & 0.7489 & 23.4794 & 0.8179 & 0.1879 \\
 & \cellcolor{gray!20}CODA & \cellcolor{gray!20}0.7815 & \cellcolor{gray!20}26.8527 & \cellcolor{gray!20}0.8768 & \cellcolor{gray!20}0.1016 \\
\midrule
\multirow{4}{*}{\shortstack{CogVideoX \cite{yang_cogvideox_2025}}} & Original & 0.7744 & - & - & - \\
 & PAB-Cons & 0.7733 & 30.8723 & 0.9365 & 0.0578 \\
 & PAB-Aggr & 0.7630 & 26.4205 & 0.8882 & 0.1135 \\
 & \cellcolor{gray!20}CODA & \cellcolor{gray!20}0.7642 & \cellcolor{gray!20}27.8768 & \cellcolor{gray!20}0.9078 & \cellcolor{gray!20}0.0905 \\
\midrule
\multirow{4}{*}{\shortstack{Open-Sora  Plan~\cite{opensora-plan}}} & Original & 0.8070 & - & - & - \\
 & PAB-Cons & 0.7825 & 20.3832 & 0.6703 & 0.3145 \\
 & PAB-Aggr & 0.6685 & 16.8197 & 0.4730 & 0.5436 \\
 & \cellcolor{gray!20}CODA & \cellcolor{gray!20}0.7676 & \cellcolor{gray!20}17.1897 & \cellcolor{gray!20}0.4722 & \cellcolor{gray!20}0.4965 \\
\midrule
\multirow{4}{*}{\shortstack{Vchitect-2.0 \cite{fan_vchitect-20_2025}}} & Original & 0.7944 & - & - & - \\
 & PAB-Cons & 0.7941 & 28.1246 & 0.8945 & 0.0407 \\
 & PAB-Aggr & 0.7907 & 27.3833 & 0.8884 & 0.0498 \\
 & \cellcolor{gray!20}CODA & \cellcolor{gray!20}0.7923 & \cellcolor{gray!20}28.0815 & \cellcolor{gray!20}0.8958 & \cellcolor{gray!20}0.0409 \\
\midrule
\multirow{4}{*}{Wan 2.1 \cite{wan_wan_2025}} & Original & 0.7434 & - & - & - \\
 & PAB-Cons & 0.7316 & 26.3462 & 0.8535 & 0.0836 \\
 & PAB-Aggr & 0.7096 & 21.6930 & 0.7397 & 0.1986 \\
 & \cellcolor{gray!20}CODA & \cellcolor{gray!20}0.7277 & \cellcolor{gray!20}24.3621 & \cellcolor{gray!20}0.7978 & \cellcolor{gray!20}0.1483 \\
\midrule
\multirow{4}{*}{HunyuanVideo \cite{kong2025hunyuanvideo}} & Original & 0.8120 & - & - & - \\
 & PAB-Cons & 0.8013 & 27.9286 & 0.8793 & 0.1082 \\
 & PAB-Aggr & 0.7853 & 26.8234 & 0.8640 & 0.1316 \\
 & \cellcolor{gray!20}CODA & \cellcolor{gray!20}0.7939 & \cellcolor{gray!20}27.6777 & \cellcolor{gray!20}0.8749 & \cellcolor{gray!20}0.1152 \\
\bottomrule
\end{tabular}
}
\caption{Quality evaluation. PAB-Cons and PAB-Aggr denote the conservative and aggressive configurations of PAB~\cite{zhao_pab_2025}, with specific parameters varying by model. PSNR, SSIM, and LPIPS are calculated against the original model results.}
\label{tab:eval_summary}
\end{table}

\subsection{Video Generation Fidelity}
\label{subsec:generation_fidelity}
As shown in Table~\ref{tab:eval_summary}, CODA remains close to the original models and to conservative PAB in overall quality. Across all evaluated models, its VBench scores stay close to the original results. More importantly, compared with PAB-Aggr, CODA generally preserves better quality across the reported metrics at a similar quality-performance operating point, indicating that its system-level gains do not come at the cost of more severe quality degradation. Overall, CODA achieves a more favorable quality-performance trade-off.

For some models, such as Latte and Open-Sora, CODA is slightly worse than PAB-Cons in PSNR, SSIM, or LPIPS, indicating larger low-level reconstruction error. However, its VBench score is slightly higher. This suggests that low-level frame-wise distortion metrics and perceptual video quality do not always vary in exactly the same direction. Taken together, these results indicate that CODA remains close to the quality level of conservative PAB while avoiding the more noticeable degradation introduced by aggressive caching.

\subsection{Ablation Study and Latency Breakdown} \label{subsec:ablation}

\begin{figure}[t]
    \centering
    \includegraphics[width=0.985\linewidth]{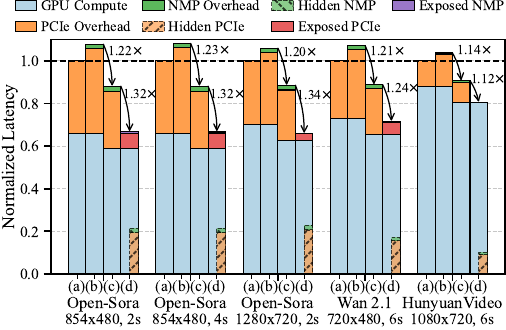}
    \caption{Ablation study and latency breakdown: (a) Cache-HostOffload, (b) +Lightweight DIMM-NMP (naive), (c) +Hybrid Hardware-Aware Caching Scheduler, (d) +CFG-Interleaved Pipelining (full CODA). Bars further separate exposed and hidden PCIe/NMP overhead to illustrate the effects of scheduling and overlap.} 
    \label{fig:ablation}
\end{figure}

While Figure~\ref{fig:speedup} reports the full-stack end-to-end gain at a similar video-quality operating point, Figure~\ref{fig:ablation} further reveals how CODA’s system mechanisms translate cache-side reuse into actual latency reduction. Specifically, it separates the overall gain into two consecutive stages: the reorganization gain from Figure~\ref{fig:ablation}(b) to (c), and the additional latency hiding from Figure~\ref{fig:ablation}(c) to (d).

First, adding Lightweight DIMM-NMP alone does not yield visible latency reduction and instead exposes additional NMP-side overhead, showing that near-memory execution capability alone cannot automatically convert cache reuse into end-to-end speedup. Nevertheless, Lightweight DIMM-NMP remains necessary because it lets cache operators complete independently on the memory side, establishing physical separation between the compute and cache paths for subsequent optimization.

The scheduler then produces the first visible latency reduction from Figure~\ref{fig:ablation}(b) to (c) by reorganizing fragmented cache activity into hardware-friendly coalesced segments. This reduces exposed PCIe and NMP overhead from fine-grained, highly interleaved triggering, showing that system-level scheduling is needed to turn disaggregation into a measurable end-to-end benefit.

CFG-Interleaved Pipelining provides the second stage of gain from Figure~\ref{fig:ablation}(c) to (d). After scheduling reorganizes the cache path, pipelining uses the GPU compute window of the other CFG branch to hide remaining PCIe/NMP overhead and shorten the critical path. This effect becomes stronger under heavier workloads because longer GPU operators can absorb more communication and near-memory execution time. For HunyuanVideo, GPU computation already dominates the timeline, so cache-side overhead is almost entirely hidden, while the end-to-end gain remains bounded by the small non-compute portion left to optimize. 


\begin{figure}[t]
    \centering
    \includegraphics[width=\linewidth]{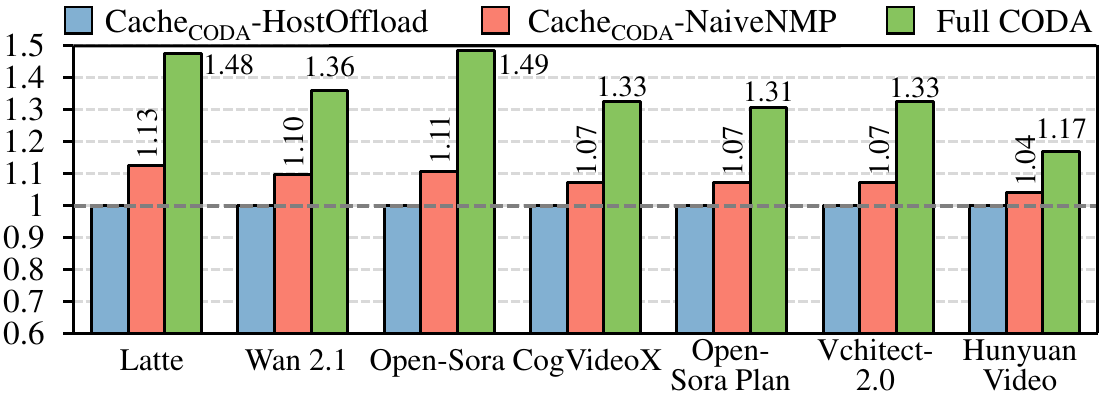}
    \caption{\revision{Ablation study with same CODA caching policy: Cache$_{\mathrm{CODA}}$ denotes variants that use the cache trace produced by CODA's Hybrid Hardware-Aware Caching Scheduler.}}
    \label{fig:matched_ablation}
\end{figure}

\phantomsection\label{rev:matched_scheduler_ablation}
\revision{To isolate the execution stack under the same CODA caching policy, we add a matched-scheduler ablation and report the largest evaluated resolution/length setting for each model. With the CODA caching policy, NaiveNMP already outperforms the matched HostOffload baseline, showing that the scheduler-generated cache trace is more amenable to NMP execution. Full CODA still achieves higher gains because CFG-interleaved pipelining overlaps the NMP-enabled independent cache path with xPU computation.}

\subsection{GPU Utilization Across Denoising Timesteps}

\begin{figure}[t]
    \centering
    \includegraphics[width=0.99\linewidth]{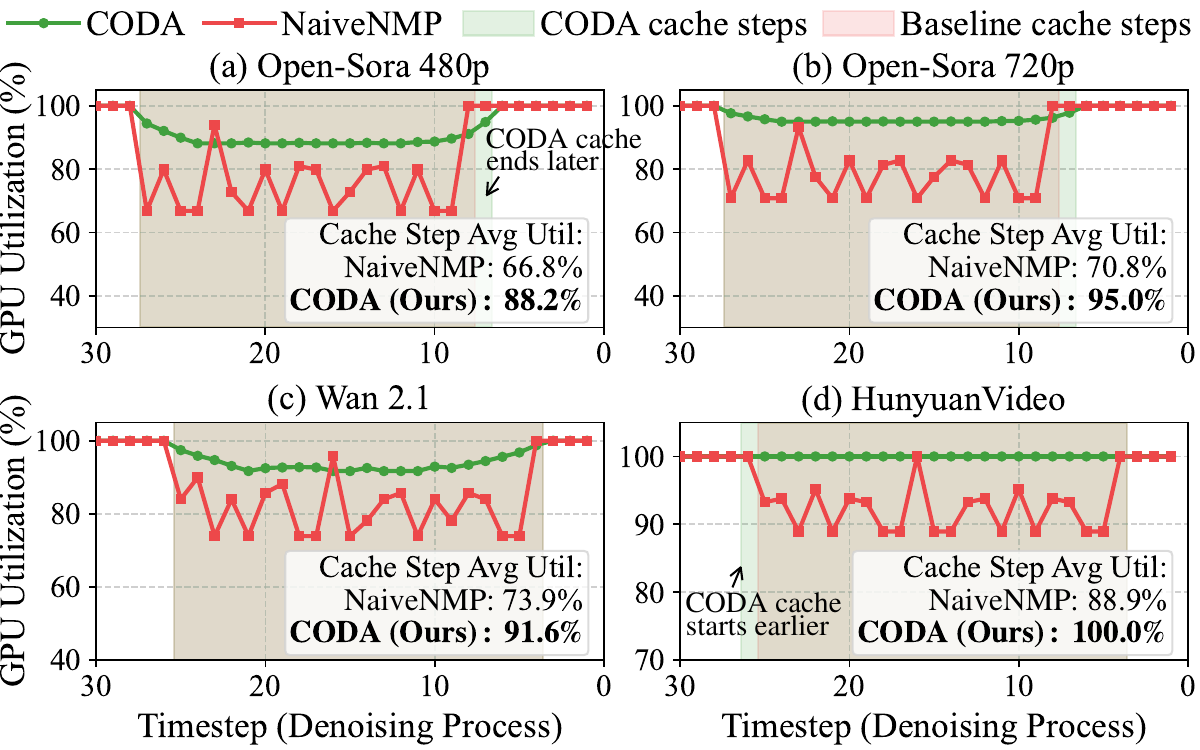}
    \caption{GPU dense-compute utilization across timesteps for CODA and Cache-NaiveNMP. Shaded regions indicate cache-step intervals for CODA and baseline (default PAB).}
    
    \label{fig:utilization}
\end{figure}
Figure~\ref{fig:utilization} compares CODA and Cache-NaiveNMP (NaiveNMP in the figure) using GPU dense-compute utilization across timesteps, normalized to non-cache timesteps (100\%).

NaiveNMP shows lower and more fluctuating utilization during cache steps, especially within the default-PAB intervals shaded red. This indicates that cache-side execution and communication remain exposed on the critical path, interrupting continuous GPU computation. In contrast, CODA keeps utilization higher and steadier by reducing cache-side interference and keeping the GPU active in dense computation.
The figure also shows that heavier workloads create larger hiding windows because dominant GPU operators run longer. CODA therefore generally maintains higher normalized utilization in these settings. For some workloads, CODA also applies caching over slightly wider timestep ranges while preserving higher utilization and competitive generation quality. 

\subsection{Efficacy of Dynamic Runtime Adjustment}

\begin{figure}[t]
    \centering
    \includegraphics[width=\linewidth]{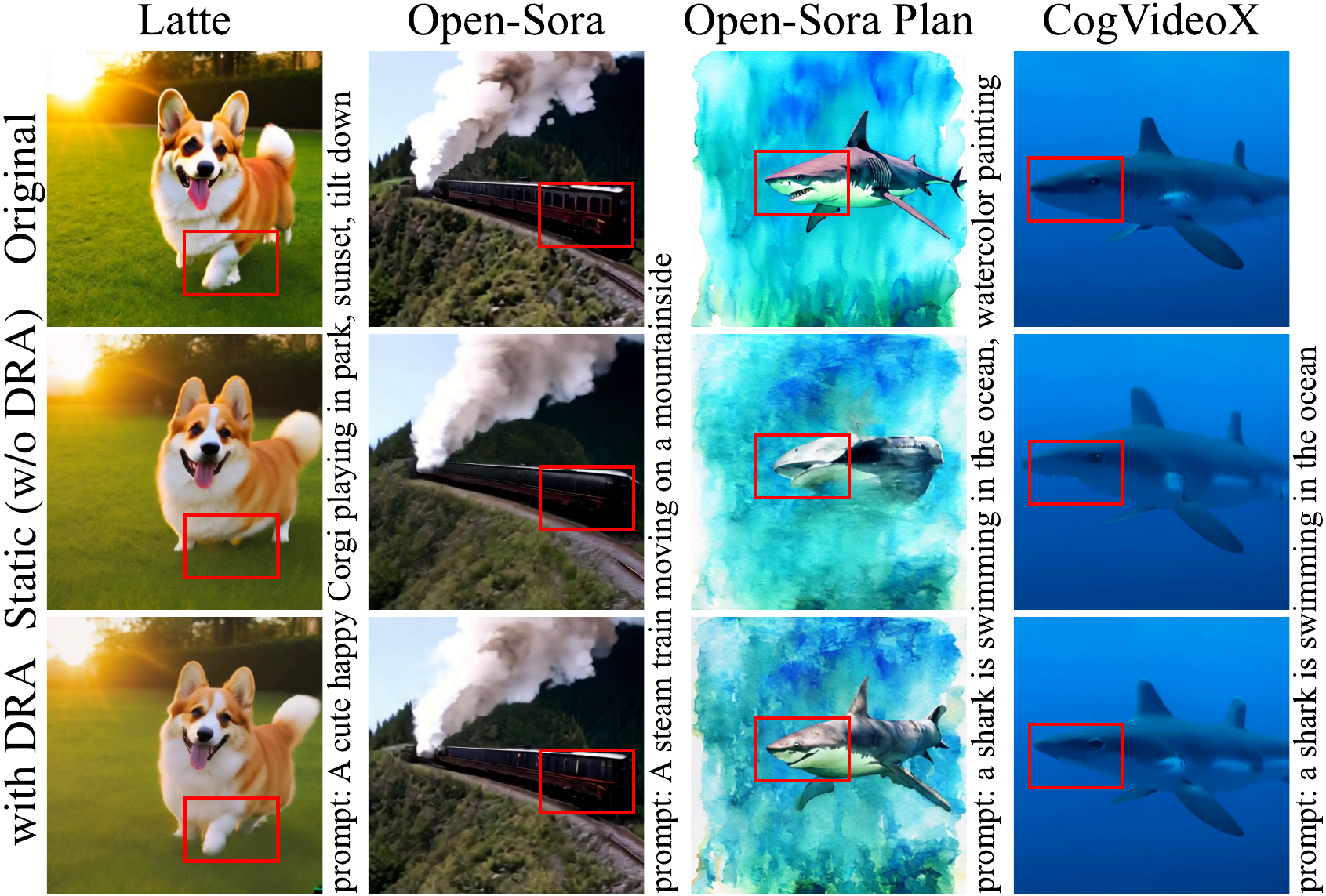}
    \caption{DRA restores visible local details of corner cases. Red boxes highlight the local details recovered by DRA.}
    \label{fig:Corner cases}
\end{figure}

Although CODA's static caching policy preserves generation quality well in most cases, a small number of prompts can still trigger DRA. DRA then conservatively adjusts the planned cache-entry behavior near the start of the caching phase. Across all evaluated models, the triggered fraction remains below 7\%. Figure~\ref{fig:Corner cases} shows representative cases where visible local degradation appears without DRA, while enabling DRA restores local details. Thus, DRA can mitigate visible local artifacts despite being activated only rarely.

\begin{figure}[t]
    \centering
    \includegraphics[width=\linewidth]{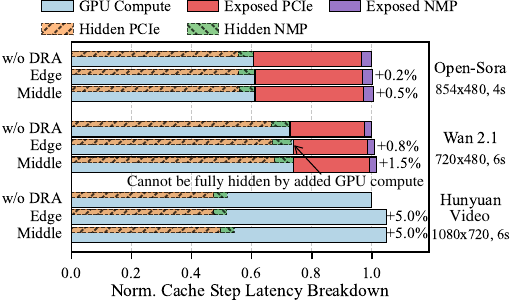}
    \caption{Latency breakdown at the triggered timestep when an operator falls back to conservative mode under DRA.}
    \label{fig:dra_overhead}
\end{figure}

Figure~\ref{fig:dra_overhead} quantifies DRA's cost by showing the normalized cache-step latency when one operator falls back to conservative mode, with annotations marking the incremental overhead at that timestep. The main added latency comes from replacing a cache operator with GPU compute, while shadow monitoring itself adds no exposed critical-path latency. \textit{Edge} and \textit{Middle} denote whether the fallback operator lies at the boundary or inside the original coarse-grained coalesced segment. An edge fallback introduces no extra PCIe latency, and the added GPU compute can still cover part of the original PCIe communication and NMP execution. A middle fallback introduces extra PCIe overhead, but GPU compute can hide part of that cost. Because such fallbacks occur only at a small number of triggered timesteps and are amortized over the denoising process, their end-to-end impact remains small. Thus, DRA provides a low-cost runtime safeguard for rare corner cases.

\subsection{Architecture DSE and Hardware Overhead}
\label{subsec:architecture_dse}

\begin{figure}[t]
    \centering
    \includegraphics[width=0.99\linewidth]{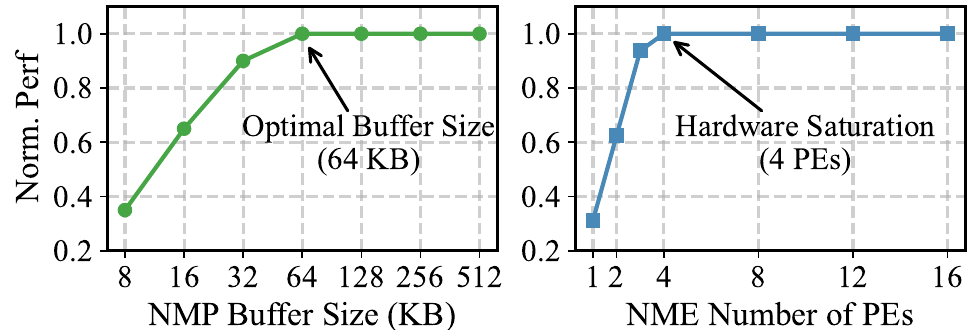}
    \caption{Architecture DSE for NMP buffer and PE numbers.} 
    \label{fig:dse}
\end{figure}

Figure~\ref{fig:dse} explores the proposed NMP buffer size and PE count. Performance improves with larger buffers but largely saturates around 64~KB, and increases up to 4 PEs before the benefit becomes limited. We therefore adopt a lightweight 64~KB buffer, 4-PE, 1~GHz configuration that captures most of CODA's system-level benefit.
We synthesize the NMP engine RTL with Synopsys Design Compiler in TSMC 7nm and obtain 64~KB SRAM parameters from a commercial SRAM compiler. A single NME occupies about 0.029~mm$^2$ and consumes about 10.17mW. Compared with a 13W DIMM~\cite{kwon_tensordimm_2019} and a typical 100~mm$^2$ memory buffer chip~\cite{DIMM-Area}, both area and power overheads are below 0.1\%, showing that the lightweight NMP configuration is sufficient for CODA's system-level gains.

\section{\revision{Related Work}} \label{sec:related_work}
\phantomsection\label{rev:related_ctc}
\noindent\textbf{\revision{Cross-timestep caching.}}
\revision{Existing CTC methods exploit cross-timestep similarity to construct cache policies at different granularities, from fixed reuse rules over timesteps or operator types, as in FORA~\cite{selvaraju_fora_2024} and PAB~\cite{zhao_pab_2025}, to token-wise caching~\cite{ToCa2025_localfix}, timestep-aware estimation~\cite{fan_taocache_2025} and forecasting-based reuse~\cite{liu_reusing_2025}. These methods focus on reusing features, operators, or tokens to reduce redundant computation while controlling quality. CODA builds on this foundation but formulates cache reuse as hardware-aware planning: it uses cacheability as a scheduling signal and combines it with communication and execution costs in the cache-policy search.}

\phantomsection\label{rev:related_vdm_accelerators}
\noindent\textbf{\revision{Video diffusion acceleration.}}
\revision{Another line of work accelerates native VDM computation within existing operators. Sparse VideoGen \cite{SparseVideoGen} exploits spatial-temporal attention sparsity, while ViDA~\cite{ding_vida_2025} and FlightVGM~\cite{FlightVGM} use differential approximation, activation sparsification, hybrid precision, and accelerator dataflow. These methods usually reduce the computation or execution cost within existing dense operators. CODA is complementary: after CTC replaces selected dense operators with cache operators, CODA focuses on executing this cache path under overflow through DIMM-side execution, coalesced scheduling, and CFG-overlapped execution.}

\section{Discussion and Future Work}
\label{sec:discussion}

The current Hybrid Hardware-Aware Caching Scheduler is one instantiation of a broader hardware-aware scheduling paradigm, rather than a fixed policy tied to current VDMs and CTC settings. In CODA, the scheduler jointly considers model-side cacheability and system-side execution cost to decide where caching should be applied, where conservative execution should be retained, and how fragmented cache operators should be reorganized into hardware-friendly segments. Thus, the scheduler is not bound to the current profiling metric or cache policy. As VDM architectures evolve and future CTC algorithms introduce new reuse opportunities and quality-performance tradeoffs, the framework can absorb updated stability signals, risk estimators, and cost terms while preserving the compute-cache disaggregation architecture.

Another direction is to extend CODA beyond VDMs to diffusion language models (DLLMs)~\cite{nie2025LLDM,chen2026dpad,cheng2025sdarsynergisticdiffusionautoregressionparadigm,bie2025llada20scalingdiffusionlanguage}. Recent DLLM studies exploit temporal stability across denoising steps through KV-cache reuse and selective recomputation, suggesting that diffusion-based text generation may also exhibit interplay between newly computed and reused cached states~\cite{wu2026fastdllm,ma2025dkvcache,jiang2026dcache}. This makes compute-cache disaggregation potentially useful for DLLM acceleration. However, this extension would not directly port the current CODA pipeline. Compared with VDMs, DLLMs expose different token-level dynamics, different cache objects, and may not provide the CFG-based overlap opportunity. A DLLM-oriented CODA would likely require redesigned scheduling signals and a new overlap mechanism tailored to token/KV evolution, which we leave for future work.

\section{Conclusion}
\label{sec:conclusion}
This paper presents \name, an algorithm-hardware co-designed architecture to address the system bottlenecks of edge video diffusion generation. We show that since cached states can easily exceed on-device VRAM and be forced into host memory, the key limitation is no longer only redundant computation, but also the PCIe communication and serialization overhead introduced by the cache path. \name therefore centers on compute-cache operator disaggregation: dense compute remains on the xPU, while memory-bound cache operators are offloaded to a lightweight DIMM-NMP subsystem. Combined with a hardware-aware caching scheduler and CFG-Interleaved Pipelining, this design transforms fragmented, blocking cache execution into a reorganized and overlapped execution flow. Experimental results show that \name converts the potential benefit of cross-timestep caching into real end-to-end system gains, achieving up to 1.80$\times$ speedup and 1.74$\times$ higher energy efficiency with high generation quality and low hardware overhead. These results suggest that, for edge VDM deployment, the key is not merely to move cache-related work closer to memory, but to co-design operator disaggregation, scheduling, and execution overlap around the communication and serialization bottlenecks.


\newpage
\bibliographystyle{ACM-Reference-Format-local-numeric}
\bibliography{ref}

@article{ma_latte_2024,
    title={Latte: Latent Diffusion Transformer for Video Generation},
    author={Ma, Xin and Wang, Yaohui and Chen, Xinyuan and Jia, Gengyun and Liu, Ziwei and Li, Yuan-Fang and Chen, Cunjian and Qiao, Yu},
    journal={Transactions on Machine Learning Research},
    year={2025}
}

@article{opensora-plan,
  title={Open-Sora Plan: Open-Source Large Video Generation Model},
  author={Lin, Bin and Ge, Yunyang and Cheng, Xinhua and Li, Zongjian and Zhu, Bin and Wang, Shaodong and He, Xianyi and Ye, Yang and Yuan, Shenghai and Chen, Liuhan and others},
  journal={arXiv preprint arXiv:2412.00131},
  year={2024}
}

@misc{kong2025hunyuanvideo,
      title={HunyuanVideo: A Systematic Framework For Large Video Generative Models}, 
      author={Weijie Kong and Qi Tian and Zijian Zhang and Rox Min and Zuozhuo Dai and Jin Zhou and Jiangfeng Xiong and Xin Li and Bo Wu and Jianwei Zhang and Kathrina Wu and Qin Lin and Junkun Yuan and Yanxin Long and Aladdin Wang and Andong Wang and Changlin Li and Duojun Huang and Fang Yang and Hao Tan and Hongmei Wang and Jacob Song and Jiawang Bai and Jianbing Wu and Jinbao Xue and Joey Wang and Kai Wang and Mengyang Liu and Pengyu Li and Shuai Li and Weiyan Wang and Wenqing Yu and Xinchi Deng and Yang Li and Yi Chen and Yutao Cui and Yuanbo Peng and Zhentao Yu and Zhiyu He and Zhiyong Xu and Zixiang Zhou and Zunnan Xu and Yangyu Tao and Qinglin Lu and Songtao Liu and Dax Zhou and Hongfa Wang and Yong Yang and Di Wang and Yuhong Liu and Jie Jiang and Caesar Zhong},
      year={2025},
      eprint={2412.03603},
      archivePrefix={arXiv},
      primaryClass={cs.CV},
      url={https://arxiv.org/abs/2412.03603}, 
}

@article{SparseVideoGen,
  title={Sparse VideoGen: Accelerating Video Diffusion Transformers with Spatial-Temporal Sparsity},
  author={Xi, Haocheng and Yang, Shuo and Zhao, Yilong and Xu, Chenfeng and Li, Muyang and Li, Xiuyu and Lin, Yujun and Cai, Han and Zhang, Jintao and Li, Dacheng and others},
  journal={International Conference on Machine Learning},
  year={2025}
}

@inproceedings{FlightVGM,
    author = {Liu, Jun and Zeng, Shulin and Ding, Li and Soedarmadji, Widyadewi and Zhou, Hao and Wang, Zehao and Li, Jinhao and Li, Jintao and Dai, Yadong and Wen, Kairui and He, Shan and Sun, Yaqi and Wang, Yu and Dai, Guohao},
    title = {FlightVGM: Efficient Video Generation Model Inference with Online Sparsification and Hybrid Precision on FPGAs},
    year = {2025},
    isbn = {9798400713965},
    publisher = {Association for Computing Machinery},
    address = {New York, NY, USA},
    url = {https://doi.org/10.1145/3706628.3708864},
    doi = {10.1145/3706628.3708864},
    booktitle = {Proceedings of the 2025 ACM/SIGDA International Symposium on Field Programmable Gate Arrays},
    pages = {2–13},
    numpages = {12},
    location = {Monterey, CA, USA},
    series = {FPGA '25}
}

@inproceedings{efficientdm-quant,
title={Efficient{DM}: Efficient Quantization-Aware Fine-Tuning of Low-Bit Diffusion Models},
author={Yefei He and Jing Liu and Weijia Wu and Hong Zhou and Bohan Zhuang},
booktitle={The Twelfth International Conference on Learning Representations},
year={2024},
url={https://openreview.net/forum?id=UmMa3UNDAz}
}

@misc{laptop-pruning,
      title={LAPTOP-Diff: Layer Pruning and Normalized Distillation for Compressing Diffusion Models}, 
      author={Dingkun Zhang and Sijia Li and Chen Chen and Qingsong Xie and Haonan Lu},
      year={2025},
      eprint={2404.11098},
      archivePrefix={arXiv},
      primaryClass={cs.CV},
      url={https://arxiv.org/abs/2404.11098}, 
}

@misc{efficientdit-distillation,
      title={Efficient-vDiT: Efficient Video Diffusion Transformers With Attention Tile}, 
      author={Hangliang Ding and Dacheng Li and Runlong Su and Peiyuan Zhang and Zhijie Deng and Ion Stoica and Hao Zhang},
      year={2025},
      eprint={2502.06155},
      archivePrefix={arXiv},
      primaryClass={cs.CV},
      url={https://arxiv.org/abs/2502.06155}, 
}

@misc{ToCa2025_localfix,
      title={Accelerating Diffusion Transformers with Token-wise Feature Caching}, 
      author={Chang Zou and Xuyang Liu and Ting Liu and Siteng Huang and Linfeng Zhang},
      year={2025},
      eprint={2410.05317},
      archivePrefix={arXiv},
      primaryClass={cs.LG},
      url={https://arxiv.org/abs/2410.05317}, 
}

@book{SA,
  title={Simulated annealing},
  author={Van Laarhoven, Peter JM and Aarts, Emile HL and van Laarhoven, Peter JM and Aarts, Emile HL},
  year={1987},
  publisher={Springer}
}

@misc{DiT2023,
      title={Scalable Diffusion Models with Transformers}, 
      author={William Peebles and Saining Xie},
      year={2023},
      eprint={2212.09748},
      archivePrefix={arXiv},
      primaryClass={cs.CV},
      url={https://arxiv.org/abs/2212.09748}, 
}

@misc{yang_cogvideox_2025,
	title = {{CogVideoX}: {Text}-to-{Video} {Diffusion} {Models} with {An} {Expert} {Transformer}},
	shorttitle = {{CogVideoX}},
	url = {http://arxiv.org/abs/2408.06072},
	doi = {10.48550/arXiv.2408.06072},
	urldate = {2025-04-13},
	publisher = {arXiv},
	author = {Yang, Zhuoyi and Teng, Jiayan and Zheng, Wendi and Ding, Ming and Huang, Shiyu and Xu, Jiazheng and Yang, Yuanming and Hong, Wenyi and Zhang, Xiaohan and Feng, Guanyu and Yin, Da and Zhang, Yuxuan and Wang, Weihan and Cheng, Yean and Xu, Bin and Gu, Xiaotao and Dong, Yuxiao and Tang, Jie},
	month = mar,
	year = {2025},
	note = {arXiv:2408.06072 [cs]},
}

@misc{blattmann_stable_2023,
	title = {Stable {Video} {Diffusion}: {Scaling} {Latent} {Video} {Diffusion} {Models} to {Large} {Datasets}},
	shorttitle = {Stable {Video} {Diffusion}},
	url = {http://arxiv.org/abs/2311.15127},
	doi = {10.48550/arXiv.2311.15127},
	urldate = {2025-04-13},
	publisher = {arXiv},
	author = {Blattmann, Andreas and Dockhorn, Tim and Kulal, Sumith and Mendelevitch, Daniel and Kilian, Maciej and Lorenz, Dominik and Levi, Yam and English, Zion and Voleti, Vikram and Letts, Adam and Jampani, Varun and Rombach, Robin},
	month = nov,
	year = {2023},
	note = {arXiv:2311.15127 [cs]},
}

@misc{selvaraju_fora_2024,
	title = {{FORA}: {Fast}-{Forward} {Caching} in {Diffusion} {Transformer} {Acceleration}},
	shorttitle = {{FORA}},
	url = {http://arxiv.org/abs/2407.01425},
	doi = {10.48550/arXiv.2407.01425},
	language = {en},
	urldate = {2025-08-06},
	publisher = {arXiv},
	author = {Selvaraju, Pratheba and Ding, Tianyu and Chen, Tianyi and Zharkov, Ilya and Liang, Luming},
	month = jul,
	year = {2024},
	note = {arXiv:2407.01425 [cs]},
}

@misc{wan_wan_2025,
	title = {Wan: {Open} and {Advanced} {Large}-{Scale} {Video} {Generative} {Models}},
	shorttitle = {Wan},
	url = {http://arxiv.org/abs/2503.20314},
	doi = {10.48550/arXiv.2503.20314},
	urldate = {2025-08-09},
	publisher = {arXiv},
	author = {Wan, Team and Wang, Ang and Ai, Baole and Wen, Bin and Mao, Chaojie and Xie, Chen-Wei and Chen, Di and Yu, Feiwu and Zhao, Haiming and Yang, Jianxiao and Zeng, Jianyuan and Wang, Jiayu and Zhang, Jingfeng and Zhou, Jingren and Wang, Jinkai and Chen, Jixuan and Zhu, Kai and Zhao, Kang and Yan, Keyu and Huang, Lianghua and Feng, Mengyang and Zhang, Ningyi and Li, Pandeng and Wu, Pingyu and Chu, Ruihang and Feng, Ruili and Zhang, Shiwei and Sun, Siyang and Fang, Tao and Wang, Tianxing and Gui, Tianyi and Weng, Tingyu and Shen, Tong and Lin, Wei and Wang, Wei and Wang, Wei and Zhou, Wenmeng and Wang, Wente and Shen, Wenting and Yu, Wenyuan and Shi, Xianzhong and Huang, Xiaoming and Xu, Xin and Kou, Yan and Lv, Yangyu and Li, Yifei and Liu, Yijing and Wang, Yiming and Zhang, Yingya and Huang, Yitong and Li, Yong and Wu, You and Liu, Yu and Pan, Yulin and Zheng, Yun and Hong, Yuntao and Shi, Yupeng and Feng, Yutong and Jiang, Zeyinzi and Han, Zhen and Wu, Zhi-Fan and Liu, Ziyu},
	month = apr,
	year = {2025},
	note = {arXiv:2503.20314 [cs]},
}

@inproceedings{zhao_pab_2025,
title={Real-Time Video Generation with Pyramid Attention Broadcast},
author={Xuanlei Zhao and Xiaolong Jin and Kai Wang and Yang You},
booktitle={The Thirteenth International Conference on Learning Representations},
year={2025},
url={https://openreview.net/forum?id=hDBrQ4DApF}
}

@misc{zheng_open-sora_2024,
	title = {Open-{Sora}: {Democratizing} {Efficient} {Video} {Production} for {All}},
	shorttitle = {Open-{Sora}},
	url = {http://arxiv.org/abs/2412.20404},
	doi = {10.48550/arXiv.2412.20404},
	language = {en},
	urldate = {2025-10-11},
	publisher = {arXiv},
	author = {Zheng, Zangwei and Peng, Xiangyu and Yang, Tianji and Shen, Chenhui and Li, Shenggui and Liu, Hongxin and Zhou, Yukun and Li, Tianyi and You, Yang},
	month = dec,
	year = {2024},
	note = {arXiv:2412.20404 [cs]},
}

@misc{fan_vchitect-20_2025,
	title = {Vchitect-2.0: {Parallel} {Transformer} for {Scaling} {Up} {Video} {Diffusion} {Models}},
	shorttitle = {Vchitect-2.0},
	url = {http://arxiv.org/abs/2501.08453},
	doi = {10.48550/arXiv.2501.08453},
	urldate = {2026-03-16},
	publisher = {arXiv},
	author = {Fan, Weichen and Si, Chenyang and Song, Junhao and Yang, Zhenyu and He, Yinan and Zhuo, Long and Huang, Ziqi and Dong, Ziyue and He, Jingwen and Pan, Dongwei and Wang, Yi and Jiang, Yuming and Wang, Yaohui and Gao, Peng and Chen, Xinyuan and Li, Hengjie and Lin, Dahua and Qiao, Yu and Liu, Ziwei},
	month = jan,
	year = {2025},
	note = {arXiv:2501.08453 [cs]},
}

@inproceedings{noauthor_mobilediffusion_nodate,
author = {Zhao, Yang and Xu, Yanwu and Xiao, Zhisheng and Jia, Haolin and Hou, Tingbo},
title = {MobileDiffusion: Instant Text-to-Image Generation on~Mobile Devices},
year = {2024},
isbn = {978-3-031-73032-0},
publisher = {Springer-Verlag},
address = {Berlin, Heidelberg},
url = {https://doi.org/10.1007/978-3-031-73033-7_13},
doi = {10.1007/978-3-031-73033-7_13},
booktitle = {Computer Vision -- ECCV 2024: 18th European Conference, Milan, Italy, September 29--October 4, 2024, Proceedings, Part LXII},
pages = {225--242},
numpages = {18},
location = {Milan, Italy}
}

@article{li_edge_2020,
	title = {Edge {AI}: {On}-{Demand} {Accelerating} {Deep} {Neural} {Network} {Inference} via {Edge} {Computing}},
	volume = {19},
	issn = {1558-2248},
	shorttitle = {Edge {AI}},
	url = {https://ieeexplore.ieee.org/document/8876870/},
	doi = {10.1109/TWC.2019.2946140},
	number = {1},
	urldate = {2026-03-16},
	journal = {IEEE Transactions on Wireless Communications},
	author = {Li, En and Zeng, Liekang and Zhou, Zhi and Chen, Xu},
	month = jan,
	year = {2020},
	pages = {447--457},
}

@misc{huang_vbench_2023,
	title = {{VBench}: {Comprehensive} {Benchmark} {Suite} for {Video} {Generative} {Models}},
	shorttitle = {{VBench}},
	url = {http://arxiv.org/abs/2311.17982},
	doi = {10.48550/arXiv.2311.17982},
	urldate = {2026-03-17},
	publisher = {arXiv},
	author = {Huang, Ziqi and He, Yinan and Yu, Jiashuo and Zhang, Fan and Si, Chenyang and Jiang, Yuming and Zhang, Yuanhan and Wu, Tianxing and Jin, Qingyang and Chanpaisit, Nattapol and Wang, Yaohui and Chen, Xinyuan and Wang, Limin and Lin, Dahua and Qiao, Yu and Liu, Ziwei},
	month = nov,
	year = {2023},
	note = {arXiv:2311.17982 [cs]},
}

@misc{xing_survey_2024,
	title = {A {Survey} on {Video} {Diffusion} {Models}},
	url = {http://arxiv.org/abs/2310.10647},
	doi = {10.48550/arXiv.2310.10647},
	urldate = {2026-03-17},
	publisher = {arXiv},
	author = {Xing, Zhen and Feng, Qijun and Chen, Haoran and Dai, Qi and Hu, Han and Xu, Hang and Wu, Zuxuan and Jiang, Yu-Gang},
	month = sep,
	year = {2024},
	note = {arXiv:2310.10647 [cs]},
}

@misc{zheng_diffusion_2025,
	title = {Diffusion {Models} on the {Edge}: {Challenges}, {Optimizations}, and {Applications}},
	shorttitle = {Diffusion {Models} on the {Edge}},
	url = {http://arxiv.org/abs/2504.15298},
	doi = {10.48550/arXiv.2504.15298},
	urldate = {2026-03-17},
	publisher = {arXiv},
	author = {Zheng, Dongqi},
	month = apr,
	year = {2025},
	note = {arXiv:2504.15298 [cs]},
}

@inproceedings{kanipakam_privacy-preserving_2025,
	title = {Privacy-{Preserving} {AI} {Inference} in {Edge} {Systems}: {Ethical} and {Architectural} {Tradeoffs}},
	issn = {2374-9628},
	shorttitle = {Privacy-{Preserving} {AI} {Inference} in {Edge} {Systems}},
	url = {https://ieeexplore.ieee.org/abstract/document/11304649},
	doi = {10.1109/IPCCC66453.2025.11304649},
	urldate = {2026-03-17},
	booktitle = {2025 {IEEE} {International} {Performance}, {Computing}, and {Communications} {Conference} ({IPCCC})},
	author = {Kanipakam, Sunitha and Devarajulu, Vishnupriya S and Kavarakuntla, Tulasi},
	month = nov,
	year = {2025},
	note = {ISSN: 2374-9628},
	pages = {1--6},
}

@article{asifuzzaman_survey_2023,
	title = {A survey on processing-in-memory techniques: {Advances} and challenges},
	volume = {4},
	issn = {2773-0646},
	shorttitle = {A survey on processing-in-memory techniques},
	url = {https://www.sciencedirect.com/science/article/pii/S2773064622000160},
	doi = {10.1016/j.memori.2022.100022},
	urldate = {2026-03-17},
	journal = {Memories - Materials, Devices, Circuits and Systems},
	author = {Asifuzzaman, Kazi and Miniskar, Narasinga Rao and Young, Aaron R. and Liu, Frank and Vetter, Jeffrey S.},
	month = jul,
	year = {2023},
	pages = {100022},
}

@misc{gomez-luna_benchmarking_2022,
	title = {Benchmarking a {New} {Paradigm}: {An} {Experimental} {Analysis} of a {Real} {Processing}-in-{Memory} {Architecture}},
	shorttitle = {Benchmarking a {New} {Paradigm}},
	url = {http://arxiv.org/abs/2105.03814},
	doi = {10.48550/arXiv.2105.03814},
	urldate = {2026-03-17},
	publisher = {arXiv},
	author = {Gómez-Luna, Juan and Hajj, Izzat El and Fernandez, Ivan and Giannoula, Christina and Oliveira, Geraldo F. and Mutlu, Onur},
	month = may,
	year = {2022},
	note = {arXiv:2105.03814 [cs]},
}

@inproceedings{park_attacc_2024,
	address = {New York, NY, USA},
	series = {{ASPLOS} '24},
	title = {{AttAcc}! {Unleashing} the {Power} of {PIM} for {Batched} {Transformer}-based {Generative} {Model} {Inference}},
	volume = {2},
	isbn = {979-8-4007-0385-0},
	url = {https://dl.acm.org/doi/10.1145/3620665.3640422},
	doi = {10.1145/3620665.3640422},
	urldate = {2026-03-16},
	booktitle = {Proceedings of the 29th {ACM} {International} {Conference} on {Architectural} {Support} for {Programming} {Languages} and {Operating} {Systems}, {Volume} 2},
	publisher = {Association for Computing Machinery},
	author = {Park, Jaehyun and Choi, Jaewan and Kyung, Kwanhee and Kim, Michael Jaemin and Kwon, Yongsuk and Kim, Nam Sung and Ahn, Jung Ho},
	month = apr,
	year = {2024},
	pages = {103--119},
}

@inproceedings{ke_recnmp_2020,
	title = {{RecNMP}: {Accelerating} {Personalized} {Recommendation} with {Near}-{Memory} {Processing}},
	shorttitle = {{RecNMP}},
	url = {https://ieeexplore.ieee.org/document/9138955},
	doi = {10.1109/ISCA45697.2020.00070},
	urldate = {2026-03-17},
	booktitle = {2020 {ACM}/{IEEE} 47th {Annual} {International} {Symposium} on {Computer} {Architecture} ({ISCA})},
	author = {Ke, Liu and Gupta, Udit and Cho, Benjamin Youngjae and Brooks, David and Chandra, Vikas and Diril, Utku and Firoozshahian, Amin and Hazelwood, Kim and Jia, Bill and Lee, Hsien-Hsin S. and Li, Meng and Maher, Bert and Mudigere, Dheevatsa and Naumov, Maxim and Schatz, Martin and Smelyanskiy, Mikhail and Wang, Xiaodong and Reagen, Brandon and Wu, Carole-Jean and Hempstead, Mark and Zhang, Xuan},
	month = may,
	year = {2020},
	pages = {790--803},
}

@misc{fan_taocache_2025,
	title = {{TaoCache}: {Structure}-{Maintained} {Video} {Generation} {Acceleration}},
	shorttitle = {{TaoCache}},
	url = {http://arxiv.org/abs/2508.08978},
	doi = {10.48550/arXiv.2508.08978},
	urldate = {2026-03-17},
	publisher = {arXiv},
	author = {Fan, Zhentao and Wang, Zongzuo and Zhang, Weiwei},
	month = aug,
	year = {2025},
	note = {arXiv:2508.08978 [cs]},
}

@misc{ho_classifier-free_2022,
	title = {Classifier-{Free} {Diffusion} {Guidance}},
	url = {http://arxiv.org/abs/2207.12598},
	doi = {10.48550/arXiv.2207.12598},
	urldate = {2026-03-17},
	publisher = {arXiv},
	author = {Ho, Jonathan and Salimans, Tim},
	month = jul,
	year = {2022},
	note = {arXiv:2207.12598 [cs]},
}

@misc{nava_meta-learning_2023,
	title = {Meta-{Learning} via {Classifier}(-free) {Diffusion} {Guidance}},
	url = {http://arxiv.org/abs/2210.08942},
	doi = {10.48550/arXiv.2210.08942},
	urldate = {2026-03-17},
	publisher = {arXiv},
	author = {Nava, Elvis and Kobayashi, Seijin and Yin, Yifei and Katzschmann, Robert K. and Grewe, Benjamin F.},
	month = jan,
	year = {2023},
	note = {arXiv:2210.08942 [cs]},
}

@misc{baykal_protodiffusion_2023,
	title = {{ProtoDiffusion}: {Classifier}-{Free} {Diffusion} {Guidance} with {Prototype} {Learning}},
	shorttitle = {{ProtoDiffusion}},
	url = {http://arxiv.org/abs/2307.01924},
	doi = {10.48550/arXiv.2307.01924},
	urldate = {2026-03-17},
	publisher = {arXiv},
	author = {Baykal, Gulcin and Karagoz, Halil Faruk and Binhuraib, Taha and Unal, Gozde},
	month = jul,
	year = {2023},
	note = {arXiv:2307.01924 [cs]},
}

@misc{bradley_classifier-free_2024,
	title = {Classifier-{Free} {Guidance} is a {Predictor}-{Corrector}},
	url = {http://arxiv.org/abs/2408.09000},
	doi = {10.48550/arXiv.2408.09000},
	urldate = {2026-03-17},
	publisher = {arXiv},
	author = {Bradley, Arwen and Nakkiran, Preetum},
	month = aug,
	year = {2024},
	note = {arXiv:2408.09000 [cs]},
}

@misc{chu_qncd_2024,
	title = {{QNCD}: {Quantization} {Noise} {Correction} for {Diffusion} {Models}},
	shorttitle = {{QNCD}},
	url = {http://arxiv.org/abs/2403.19140},
	doi = {10.48550/arXiv.2403.19140},
	urldate = {2026-03-17},
	publisher = {arXiv},
	author = {Chu, Huanpeng and Wu, Wei and Zang, Chengjie and Yuan, Kun},
	month = sep,
	year = {2024},
	note = {arXiv:2403.19140 [cs]},
}

@misc{relic_bridging_2025,
	title = {Bridging the {Gap} between {Gaussian} {Diffusion} {Models} and {Universal} {Quantization} for {Image} {Compression}},
	url = {http://arxiv.org/abs/2504.02579},
	doi = {10.48550/arXiv.2504.02579},
	urldate = {2026-03-17},
	publisher = {arXiv},
	author = {Relic, Lucas and Azevedo, Roberto and Zhang, Yang and Gross, Markus and Schroers, Christopher},
	month = apr,
	year = {2025},
	note = {arXiv:2504.02579 [eess]},
}

@misc{yao_timestep-aware_2024,
	title = {Timestep-{Aware} {Correction} for {Quantized} {Diffusion} {Models}},
	url = {http://arxiv.org/abs/2407.03917},
	doi = {10.48550/arXiv.2407.03917},
	urldate = {2026-03-17},
	publisher = {arXiv},
	author = {Yao, Yuzhe and Tian, Feng and Chen, Jun and Lin, Haonan and Dai, Guang and Liu, Yong and Wang, Jingdong},
	month = jul,
	year = {2024},
	note = {arXiv:2407.03917 [cs]},
}

@inproceedings{Lee_AxDIMM_2022,
author = {Lee, Donghun and So, Jinin and AHN, MINSEON and Lee, Jong-Geon and Kim, Jungmin and Cho, Jeonghyeon and Oliver, Rebholz and Thummala, Vishnu Charan and JV, Ravi shankar and Upadhya, Sachin Suresh and Khan, Mohammed Ibrahim and Kim, Jin Hyun},
title = {Improving In-Memory Database Operations with Acceleration DIMM (AxDIMM)},
year = {2022},
isbn = {9781450393782},
publisher = {Association for Computing Machinery},
address = {New York, NY, USA},
url = {https://doi.org/10.1145/3533737.3535093},
doi = {10.1145/3533737.3535093},
booktitle = {Proceedings of the 18th International Workshop on Data Management on New Hardware},
articleno = {2},
numpages = {9},
location = {Philadelphia, PA, USA},
series = {DaMoN '22}
}

@misc{videosys2024,
  author={VideoSys Team},
  title={VideoSys: An Easy and Efficient System for Video Generation},
  year={2024},
  publisher={GitHub},
  url = {https://github.com/NUS-HPC-AI-Lab/VideoSys},
}

@inproceedings{li_specpim_2024,
	address = {New York, NY, USA},
	series = {{ASPLOS} '24},
	title = {{SpecPIM}: {Accelerating} {Speculative} {Inference} on {PIM}-{Enabled} {System} via {Architecture}-{Dataflow} {Co}-{Exploration}},
	volume = {3},
	isbn = {979-8-4007-0386-7},
	shorttitle = {{SpecPIM}},
	url = {https://dl.acm.org/doi/10.1145/3620666.3651352},
	doi = {10.1145/3620666.3651352},
	urldate = {2026-03-16},
	booktitle = {Proceedings of the 29th {ACM} {International} {Conference} on {Architectural} {Support} for {Programming} {Languages} and {Operating} {Systems}, {Volume} 3},
	publisher = {Association for Computing Machinery},
	author = {Li, Cong and Zhou, Zhe and Zheng, Size and Zhang, Jiaxi and Liang, Yun and Sun, Guangyu},
	month = apr,
	year = {2024},
	pages = {950--965},
}

@misc{kwon_tensordimm_2019,
	title = {{TensorDIMM}: {A} {Practical} {Near}-{Memory} {Processing} {Architecture} for {Embeddings} and {Tensor} {Operations} in {Deep} {Learning}},
	shorttitle = {{TensorDIMM}},
	url = {http://arxiv.org/abs/1908.03072},
	doi = {10.48550/arXiv.1908.03072},
	urldate = {2026-03-17},
	publisher = {arXiv},
	author = {Kwon, Youngeun and Lee, Yunjae and Rhu, Minsoo},
	month = aug,
	year = {2019},
	note = {arXiv:1908.03072 [cs]},
}

@article{li_optimizing_2025,
	title = {Optimizing {Quantized} {Diffusion} {Models} via {Distillation} with {Cross}-{Timestep} {Error} {Correction}},
	volume = {39},
	copyright = {Copyright (c) 2025 Association for the Advancement of Artificial Intelligence},
	issn = {2374-3468},
	url = {https://ojs.aaai.org/index.php/AAAI/article/view/34039},
	doi = {10.1609/aaai.v39i17.34039},
	language = {en},
	number = {17},
	urldate = {2026-03-17},
	journal = {Proceedings of the AAAI Conference on Artificial Intelligence},
	author = {Li, Yanxi and Du, Chengbin},
	month = apr,
	year = {2025},
	pages = {18530--18538},
}

@misc{becker_edit_2025,
	title = {{EDiT}: {Efficient} {Diffusion} {Transformers} with {Linear} {Compressed} {Attention}},
	shorttitle = {{EDiT}},
	url = {http://arxiv.org/abs/2503.16726},
	doi = {10.48550/arXiv.2503.16726},
	urldate = {2026-03-17},
	publisher = {arXiv},
	author = {Becker, Philipp and Mehrotra, Abhinav and Chavhan, Ruchika and Chadwick, Malcolm and Morreale, Luca and Noroozi, Mehdi and Ramos, Alberto Gil and Bhattacharya, Sourav},
	month = aug,
	year = {2025},
	note = {arXiv:2503.16726 [cs]},
}

@misc{openai2025sora2,
  title={{Sora 2: Video Generation Models as World Simulators}},
  author={OpenAI},
  year={2026},
  howpublished={\url{https://openai.com/index/sora-2/}}
}

@misc{bytedance2025seedance,
  title={{Seedance 2.0: The Next Generation of AI Video}},
  author={ByteDance AI Lab},
  year={2026},
  howpublished={\url{https://fal.ai/seedance-2.0}}
}

@misc{ali2025wan,
  title={{Wan AI: Leading AI Video Generation Model}},
  author={Alibaba Cloud Group},
  year={2026},
  howpublished={\url{https://wan.video/}}
}

@INPROCEEDINGS{lee_aim_2022,
  author={Lee, Seongju and Kim, Kyuyoung and Oh, Sanghoon and Park, Joonhong and Hong, Gimoon and Ka, Dongyoon and Hwang, Kyudong and Park, Jeongje and Kang, Kyeongpil and Kim, Jungyeon and Jeon, Junyeol and Kim, Nahsung and Kwon, Yongkee and Vladimir, Kornijcuk and Shin, Woojae and Won, Jongsoon and Lee, Minkyu and Joo, Hyunha and Choi, Haerang and Lee, Jaewook and Ko, Donguc and Jun, Younggun and Cho, Keewon and Kim, Ilwoong and Song, Choungki and Jeong, Chunseok and Kwon, Daehan and Jang, Jieun and Park, Il and Chun, Junhyun and Cho, Joohwan},
  booktitle={2022 IEEE International Solid-State Circuits Conference (ISSCC)}, 
  title={A 1ynm 1.25V 8Gb, 16Gb/s/pin GDDR6-based Accelerator-in-Memory supporting 1TFLOPS MAC Operation and Various Activation Functions for Deep-Learning Applications}, 
  year={2022},
  volume={65},
  number={},
  pages={1-3},
  doi={10.1109/ISSCC42614.2022.9731711}}

@ARTICLE{DIMM-Area,
  author={Meaney, P. J. and Curley, L. D. and Gilda, G. D. and Hodges, M. R. and Buerkle, D. J. and Siegl, R. D. and Dong, R. K.},
  journal={IBM Journal of Research and Development}, 
  title={The IBM z13 memory subsystem for big data}, 
  year={2015},
  volume={59},
  number={4/5},
  pages={4:1-4:11},
  doi={10.1147/JRD.2015.2429031}}

@article{ramulator2,
author = {Luo, Haocong and Tu\u{g}rul, Yahya Can and Bostanc\i{}, F. Nisa and Olgun, Ataberk and Ya\u{g}l\i{}k\c{c}\i{}, A. Giray and Mutlu, Onur},
title = {Ramulator 2.0: A Modern, Modular, and Extensible DRAM Simulator},
year = {2024},
issue_date = {Jan.-June 2024},
publisher = {IEEE Computer Society},
address = {USA},
volume = {23},
number = {1},
issn = {1556-6056},
url = {https://doi.org/10.1109/LCA.2023.3333759},
doi = {10.1109/LCA.2023.3333759},
journal = {IEEE Comput. Archit. Lett.},
month = jan,
pages = {112–116},
numpages = {5}
}

@inproceedings{liu_speca_2025,
	title = {{SpeCa}: Accelerating Diffusion Transformers with Speculative Feature Caching},
	url = {http://arxiv.org/abs/2509.11628},
	doi = {10.1145/3746027.3755331},
	shorttitle = {{SpeCa}},
	pages = {10024--10033},
	booktitle = {Proceedings of the 33rd {ACM} International Conference on Multimedia},
	author = {Liu, Jiacheng and Zou, Chang and Lyu, Yuanhuiyi and Ren, Fei and Wang, Shaobo and Li, Kaixin and Zhang, Linfeng},
	urldate = {2026-03-31},
	date = {2025-10-27},
	year = {2025},
	eprinttype = {arxiv},
	eprint = {2509.11628 [cs]},
}

@misc{wu_invardiff_2025,
	title = {{InvarDiff}: Cross-Scale Invariance Caching for Accelerated Diffusion Models},
	url = {http://arxiv.org/abs/2512.05134},
	doi = {10.48550/arXiv.2512.05134},
	shorttitle = {{InvarDiff}},
	number = {{arXiv}:2512.05134},
	publisher = {{arXiv}},
	author = {Wu, Zihao},
	urldate = {2026-03-31},
	date = {2025-11-29},
	year = {2025},
	eprinttype = {arxiv},
	eprint = {2512.05134 [cs]},
}

@misc{liu_reusing_2025,
	title = {From Reusing to Forecasting: Accelerating Diffusion Models with {TaylorSeers}},
	url = {http://arxiv.org/abs/2503.06923},
	doi = {10.48550/arXiv.2503.06923},
	shorttitle = {From Reusing to Forecasting},
	number = {{arXiv}:2503.06923},
	publisher = {{arXiv}},
	author = {Liu, Jiacheng and Zou, Chang and Lyu, Yuanhuiyi and Chen, Junjie and Zhang, Linfeng},
	urldate = {2026-03-31},
	date = {2025-08-11},
	year = {2025},
	eprinttype = {arxiv},
	eprint = {2503.06923 [cs]},
}

@misc{cache-dit_2025,
  title={Cache-DiT: A PyTorch-native Inference Engine with Hybrid Cache Acceleration and Massive Parallelism for DiTs.},
  url={https://github.com/vipshop/cache-dit.git},
  author={vipshop.com, DefTruth},
  year={2025}
}

@misc{li_out_2026,
	title = {Out of the Memory Barrier: A Highly Memory Efficient Training System for {LLMs} with Million-Token Contexts},
	url = {http://arxiv.org/abs/2602.02108},
	doi = {10.48550/arXiv.2602.02108},
	shorttitle = {Out of the Memory Barrier},
	number = {{arXiv}:2602.02108},
	publisher = {{arXiv}},
	author = {Li, Wenhao and Yu, Daohai and Luo, Gen and Zhang, Yuxin and Chao, Fei and Ji, Rongrong and Wu, Yifan and Liu, Jiaxin and Gong, Ziyang and Liao, Zimu},
	urldate = {2026-03-31},
	date = {2026-02-07},
	year = {2026},
	eprinttype = {arxiv},
	eprint = {2602.02108 [cs]},
	note = {version: 2},
}

@inproceedings{maurya_mlp-offload_2025,
	location = {New York, {NY}, {USA}},
	title = {{MLP}-Offload: Multi-Level, Multi-Path Offloading for {LLM} Pre-training to Break the {GPU} Memory Wall},
	isbn = {979-8-4007-1466-5},
	url = {https://dl.acm.org/doi/10.1145/3712285.3759864},
	doi = {10.1145/3712285.3759864},
	series = {{SC} '25},
	shorttitle = {{MLP}-Offload},
	pages = {1381--1394},
	booktitle = {Proceedings of the International Conference for High Performance Computing, Networking, Storage and Analysis},
	publisher = {Association for Computing Machinery},
	author = {Maurya, Avinash Kumar and Rafique, M. Mustafa and Cappello, Franck and Nicolae, Bogdan},
	urldate = {2026-03-30},
	date = {2025-11-15},
	year = {2025},
}

@ARTICLE{LiSADIMM2025,
author={Li, Huize and Chen, Dan and Mitra, Tulika},
journal={ IEEE Transactions on Computers },
title={{ SADIMM: Accelerating $\underline{\text{S}}$parse $\underline{\text{A}}$ttention Using $\underline{\text{DIMM}}$-Based Near-Memory Processing }},
year={2025},
volume={74},
number={02},
ISSN={1557-9956},
pages={542-554},
doi={10.1109/TC.2024.3500362},
url = {https://doi.ieeecomputersociety.org/10.1109/TC.2024.3500362},
publisher={IEEE Computer Society},
address={Los Alamitos, CA, USA},
month=feb}

@article{kim_darwin_2025,
	title = {Darwin: A {DRAM}-based Multi-level Processing-in-Memory Architecture for Data Analytics},
	volume = {13},
	issn = {2168-6750, 2376-4562},
	url = {http://arxiv.org/abs/2305.13970},
	doi = {10.1109/TETC.2024.3493132},
	shorttitle = {Darwin},
	pages = {739--752},
	number = {3},
	journaltitle = {{IEEE} Transactions on Emerging Topics in Computing},
	shortjournal = {{IEEE} Trans. Emerg. Topics Comput.},
	author = {Kim, Donghyuk and Kim, Jae-Young and Han, Wontak and Won, Jongsoon and Choi, Haerang and Kwon, Yongkee and Kim, Joo-Young},
	urldate = {2026-03-31},
	date = {2025-07},
	year = {2025},
	eprinttype = {arxiv},
	eprint = {2305.13970 [eess]},
}

@inproceedings{li_ansmet_2025,
	location = {New York, {NY}, {USA}},
	title = {{ANSMET}: Approximate Nearest Neighbor Search with Near-Memory Processing and Hybrid Early Termination},
	isbn = {979-8-4007-1261-6},
	url = {https://dl.acm.org/doi/10.1145/3695053.3731013},
	doi = {10.1145/3695053.3731013},
	series = {{ISCA} '25},
	shorttitle = {{ANSMET}},
	pages = {1093--1107},
	booktitle = {Proceedings of the 52nd Annual International Symposium on Computer Architecture},
	publisher = {Association for Computing Machinery},
	author = {Li, Yiwei and Jin, Yuxin and Tian, Boyu and Zhang, Huanchen and Gao, Mingyu},
	urldate = {2026-03-30},
	date = {2025-06-20},
}

@inproceedings{gu_ipim_2020,
	title = {{iPIM}: Programmable In-Memory Image Processing Accelerator Using Near-Bank Architecture},
	url = {https://ieeexplore.ieee.org/document/9138985},
	doi = {10.1109/ISCA45697.2020.00071},
	shorttitle = {{iPIM}},
	eventtitle = {2020 {ACM}/{IEEE} 47th Annual International Symposium on Computer Architecture ({ISCA})},
	pages = {804--817},
	booktitle = {2020 {ACM}/{IEEE} 47th Annual International Symposium on Computer Architecture ({ISCA})},
	author = {Gu, Peng and Xie, Xinfeng and Ding, Yufei and Chen, Guoyang and Zhang, Weifeng and Niu, Dimin and Xie, Yuan},
	urldate = {2026-03-31},
	date = {2020-05},
	year = {2020},
}

@inproceedings{huangfu_medal_2019,
	location = {New York, {NY}, {USA}},
	title = {{MEDAL}: Scalable {DIMM} based Near Data Processing Accelerator for {DNA} Seeding Algorithm},
	isbn = {978-1-4503-6938-1},
	url = {https://dl.acm.org/doi/10.1145/3352460.3358329},
	doi = {10.1145/3352460.3358329},
	series = {{MICRO}-52},
	shorttitle = {{MEDAL}},
	pages = {587--599},
	booktitle = {Proceedings of the 52nd Annual {IEEE}/{ACM} International Symposium on Microarchitecture},
	publisher = {Association for Computing Machinery},
	author = {Huangfu, Wenqin and Li, Xueqi and Li, Shuangchen and Hu, Xing and Gu, Peng and Xie, Yuan},
	urldate = {2026-03-31},
	date = {2019-10-12},
	year = {2019},
}

@inproceedings{zhou_gnnear_2023,
	location = {New York, {NY}, {USA}},
	title = {{GNNear}: Accelerating Full-Batch Training of Graph Neural Networks with near-Memory Processing},
	isbn = {978-1-4503-9868-8},
	url = {https://dl.acm.org/doi/10.1145/3559009.3569670},
	doi = {10.1145/3559009.3569670},
	series = {{PACT} '22},
	shorttitle = {{GNNear}},
	pages = {54--68},
	booktitle = {Proceedings of the International Conference on Parallel Architectures and Compilation Techniques},
	publisher = {Association for Computing Machinery},
	author = {Zhou, Zhe and Li, Cong and Wei, Xuechao and Wang, Xiaoyang and Sun, Guangyu},
	urldate = {2026-03-31},
	date = {2023-01-27},
	year = {2023},
}

@inproceedings{liu_enmc_2021,
	location = {New York, {NY}, {USA}},
	title = {{ENMC}: Extreme Near-Memory Classification via Approximate Screening},
	isbn = {978-1-4503-8557-2},
	url = {https://dl.acm.org/doi/10.1145/3466752.3480090},
	doi = {10.1145/3466752.3480090},
	series = {{MICRO} '21},
	shorttitle = {{ENMC}},
	pages = {1309--1322},
	booktitle = {{MICRO}-54: 54th Annual {IEEE}/{ACM} International Symposium on Microarchitecture},
	publisher = {Association for Computing Machinery},
	author = {Liu, Liu and Lin, Jilan and Qu, Zheng and Ding, Yufei and Xie, Yuan},
	urldate = {2026-03-31},
	date = {2021-10-17},
	year = {2021},
}

@inproceedings{cui2026bwcache,
title={{BWC}ache: Accelerating Video Diffusion Transformers through Block-Wise Caching},
author={Hanshuai Cui and Zhiqing Tang and Zhifei Xu and Zhi Yao and Wenyi Zeng and Weijia Jia},
booktitle={The Fourteenth International Conference on Learning Representations},
year={2026},
url={https://openreview.net/forum?id=5bJZtzTFYy}
}

@misc{liu_freqca_2025,
	title = {{FreqCa}: Accelerating Diffusion Models via Frequency-Aware Caching},
	url = {http://arxiv.org/abs/2510.08669},
	doi = {10.48550/arXiv.2510.08669},
	shorttitle = {{FreqCa}},
	number = {{arXiv}:2510.08669},
	publisher = {{arXiv}},
	author = {Liu, Jiacheng and Cai, Peiliang and Zhou, Qinming and Lin, Yuqi and Kong, Deyang and Huang, Benhao and Pan, Yupei and Xu, Haowen and Zou, Chang and Tang, Junshu and Zheng, Shikang and Zhang, Linfeng},
	urldate = {2026-03-31},
	date = {2025-10-09},
	year = {2025},
	eprinttype = {arxiv},
	eprint = {2510.08669 [cs]},
	note = {version: 1},
}

@misc{liu_fastcache_2026,
	title = {{FastCache}: Fast Caching for Diffusion Transformer Through Learnable Linear Approximation},
	url = {http://arxiv.org/abs/2505.20353},
	doi = {10.48550/arXiv.2505.20353},
	shorttitle = {{FastCache}},
	number = {{arXiv}:2505.20353},
	publisher = {{arXiv}},
	author = {Liu, Dong and Yu, Yanxuan and Zhang, Jiayi and Li, Yifan and Lengerich, Ben and Wu, Ying Nian},
	urldate = {2026-03-31},
	date = {2026-03-27},
	year = {2026},
	eprinttype = {arxiv},
	eprint = {2505.20353 [cs]},
	note = {version: 3},
}

@article{ma_model_nodate2025,
    author    = {Ma, Xuran and Liu, Yexin and Liu, Yaofu and Wu, Xianfeng and Zheng, Mingzhe and Wang, Zihao and Lim, Ser-Nam and Yang, Harry},
    title     = {Model Reveals What to Cache: Profiling-Based Feature Reuse for Video Diffusion Models},
    booktitle = {Proceedings of the IEEE/CVF International Conference on Computer Vision (ICCV)},
    month     = {October},
    year      = {2025},
    pages     = {17150-17159}
}

@inproceedings{
chung2025cfg,
title={{CFG}++: Manifold-constrained Classifier Free Guidance for Diffusion Models},
author={Hyungjin Chung and Jeongsol Kim and Geon Yeong Park and Hyelin Nam and Jong Chul Ye},
booktitle={The Thirteenth International Conference on Learning Representations},
year={2025},
url={https://openreview.net/forum?id=E77uvbOTtp}
}

@inproceedings{saini2025rectified,
title={Rectified {CFG}++ for Flow Based Models},
author={Shreshth Saini and Shashank Gupta and Alan Bovik},
booktitle={The Thirty-ninth Annual Conference on Neural Information Processing Systems},
year={2025},
url={https://openreview.net/forum?id=NosdT1FHPv}
}

@article{phunyaphibarn_unconditional_nodate,
	title = {Unconditional Priors Matter! Improving Conditional Generation of Fine-Tuned Diffusion Models},
	author = {Phunyaphibarn, Prin and Lee, Phillip Y and Kim, Jaihoon and Sung, Minhyuk},
	year = {2025},
	langid = {english},
}

@online{li_towards_2025,
	title = {Towards Understanding the Mechanisms of Classifier-Free Guidance},
	url = {https://arxiv.org/abs/2505.19210v3},
	titleaddon = {{arXiv}.org},
	author = {Li, Xiang and Wang, Rongrong and Qu, Qing},
	urldate = {2026-03-31},
	date = {2025-05-25},
	year = {2025},
	langid = {english},
}

@inproceedings{liu_make_2025,
	title = {Make {LLM} Inference Affordable to Everyone: Augmenting {GPU} Memory with {NDP}-{DIMM}},
	issn = {2378-203X},
	url = {https://ieeexplore.ieee.org/document/10946712},
	doi = {10.1109/HPCA61900.2025.00129},
	shorttitle = {Make {LLM} Inference Affordable to Everyone},
	eventtitle = {2025 {IEEE} International Symposium on High Performance Computer Architecture ({HPCA})},
	pages = {1751--1765},
	booktitle = {2025 {IEEE} International Symposium on High Performance Computer Architecture ({HPCA})},
	author = {Liu, Lian and Zhao, Shixin and Li, Bing and Ren, Haimeng and Xu, Zhaohui and Wang, Mengdi and Li, Xiaowei and Han, Yinhe and Wang, Ying},
	urldate = {2025-11-07},
	date = {2025-03},
	year = {2025},
	langid = {english},
}

@inproceedings{nie2025LLDM,
title={Large Language Diffusion Models},
author={Shen Nie and Fengqi Zhu and Zebin You and Xiaolu Zhang and Jingyang Ou and Jun Hu and JUN ZHOU and Yankai Lin and Ji-Rong Wen and Chongxuan Li},
booktitle={The Thirty-ninth Annual Conference on Neural Information Processing Systems},
year={2025},
url={https://openreview.net/forum?id=KnqiC0znVF}
}

@inproceedings{chen2026dpad,
title={{DP}ad: Efficient Diffusion Language Models with Suffix Dropout},
author={Xinhua Chen and Sitao Huang and Cong Guo and Chiyue Wei and Yintao He and Jianyi Zhang and Hai Helen Li and Yiran Chen},
booktitle={The Fourteenth International Conference on Learning Representations},
year={2026},
url={https://openreview.net/forum?id=0yOsSMU1eY}
}

@inproceedings{jiang2026dcache,
title={d\${\textasciicircum}2\$Cache: Accelerating Diffusion-Based {LLM}s via Dual Adaptive Caching},
author={Yuchu Jiang and Yue Cai and Xiangzhong Luo and Jiale Fu and Jiarui Wang and Chonghan Liu and Xu Yang},
booktitle={The Fourteenth International Conference on Learning Representations},
year={2026},
url={https://openreview.net/forum?id=SjInfpK5RM}
}

@inproceedings{ma2025dkvcache,
title={d{KV}-Cache: The Cache for Diffusion Language Models},
author={Xinyin Ma and Runpeng Yu and Gongfan Fang and Xinchao Wang},
booktitle={The Thirty-ninth Annual Conference on Neural Information Processing Systems},
year={2025},
url={https://openreview.net/forum?id=Gppo2JImHs}
}

@inproceedings{wu2026fastdllm,
title={Fast-d{LLM}: Training-free Acceleration of Diffusion {LLM} by Enabling {KV} Cache and Parallel Decoding},
author={Chengyue Wu and Hao Zhang and Shuchen Xue and Zhijian Liu and Shizhe Diao and Ligeng Zhu and Ping Luo and Song Han and Enze Xie},
booktitle={The Fourteenth International Conference on Learning Representations},
year={2026},
url={https://openreview.net/forum?id=3Z3Is6hnOT}
}

@misc{cheng2025sdarsynergisticdiffusionautoregressionparadigm,
      title={SDAR: A Synergistic Diffusion-AutoRegression Paradigm for Scalable Sequence Generation}, 
      author={Shuang Cheng and Yihan Bian and Dawei Liu and Linfeng Zhang and Qian Yao and Zhongbo Tian and Wenhai Wang and Qipeng Guo and Kai Chen and Biqing Qi and Bowen Zhou},
      year={2025},
      eprint={2510.06303},
      archivePrefix={arXiv},
      primaryClass={cs.LG},
      url={https://arxiv.org/abs/2510.06303}, 
}

@misc{bie2025llada20scalingdiffusionlanguage,
      title={LLaDA2.0: Scaling Up Diffusion Language Models to 100B}, 
      author={Tiwei Bie and Maosong Cao and Kun Chen and Lun Du and Mingliang Gong and Zhuochen Gong and Yanmei Gu and Jiaqi Hu and Zenan Huang and Zhenzhong Lan and Chengxi Li and Chongxuan Li and Jianguo Li and Zehuan Li and Huabin Liu and Lin Liu and Guoshan Lu and Xiaocheng Lu and Yuxin Ma and Jianfeng Tan and Lanning Wei and Ji-Rong Wen and Yipeng Xing and Xiaolu Zhang and Junbo Zhao and Da Zheng and Jun Zhou and Junlin Zhou and Zhanchao Zhou and Liwang Zhu and Yihong Zhuang},
      year={2025},
      eprint={2512.15745},
      archivePrefix={arXiv},
      primaryClass={cs.LG},
      url={https://arxiv.org/abs/2512.15745}, 
}

@inproceedings{ding_vida_2025,
author = {Ding, Li and Liu, Jun and Huang, Shan and Dai, Guohao},
title = {ViDA: Video Diffusion Transformer Acceleration with Differential Approximation and Adaptive Dataflow},
year = {2025},
isbn = {9798400706356},
publisher = {Association for Computing Machinery},
address = {New York, NY, USA},
url = {https://doi.org/10.1145/3658617.3697692},
doi = {10.1145/3658617.3697692},
booktitle = {Proceedings of the 30th Asia and South Pacific Design Automation Conference},
pages = {148–154},
numpages = {7},
location = {Tokyo, Japan},
series = {ASPDAC '25}
}

\end{document}